\documentclass[conference, a4paper]{IEEEtran}

\pdfoutput=1

\usepackage{amsmath,bm,amssymb,amsfonts,amsthm}
\usepackage[linesnumbered, ruled, vlined]{algorithm2e}
\usepackage{graphicx}
\usepackage{xcolor} 
\usepackage{relsize}
\usepackage{textcomp}
\usepackage{listings}
\usepackage{gensymb}

\usepackage{changepage}
\usepackage{hhline}
\usepackage{multirow}
\usepackage{nomencl}
\usepackage{mathtools}
\usepackage{commath}
\usepackage{booktabs}
\usepackage{subfiles}
\usepackage{tabularx}
\usepackage{lscape}
\usepackage{rotating}

\usepackage{verbatim}
\usepackage{comment}

\usepackage[normalem]{ulem}
\usepackage[utf8]{inputenc}
\usepackage[hyphens]{url}
\usepackage{longtable}
\usepackage{lineno}
    \modulolinenumbers[5]
\usepackage{wrapfig}
\usepackage{enumitem}
    \setlist{nolistsep}
\usepackage[hidelinks]{hyperref}
\usepackage{makecell}
\usepackage{multicol}
\usepackage{colortbl}
\usepackage{subfigure}

\usepackage[left=1.57cm,right=1.57cm,top=0.95cm,bottom=2.54cm]{geometry}
\def\BibTeX{{\rm B\kern-.05em{\sc i\kern-.025em b}\kern-.08em
    T\kern-.1667em\lower.7ex\hbox{E}\kern-.125emX}}

\usepackage{scalerel}
\usepackage{tikz}
\usetikzlibrary{svg.path}
\definecolor{orcidlogocol}{HTML}{A6CE39}
\tikzset{
  orcidlogo/.pic={
    \fill[orcidlogocol] svg{M256,128c0,70.7-57.3,128-128,128C57.3,256,0,198.7,0,128C0,57.3,57.3,0,128,0C198.7,0,256,57.3,256,128z};
    \fill[white] svg{M86.3,186.2H70.9V79.1h15.4v48.4V186.2z}
                 svg{M108.9,79.1h41.6c39.6,0,57,28.3,57,53.6c0,27.5-21.5,53.6-56.8,53.6h-41.8V79.1z M124.3,172.4h24.5c34.9,0,42.9-26.5,42.9-39.7c0-21.5-13.7-39.7-43.7-39.7h-23.7V172.4z}
                 svg{M88.7,56.8c0,5.5-4.5,10.1-10.1,10.1c-5.6,0-10.1-4.6-10.1-10.1c0-5.6,4.5-10.1,10.1-10.1C84.2,46.7,88.7,51.3,88.7,56.8z};
  }
}
\newcommand\orcidicon[1]{\href{https://orcid.org/#1}{\mbox{\scalerel*{
\begin{tikzpicture}[yscale=-1,transform shape]
\pic{orcidlogo};
\end{tikzpicture}
}{|}}}}

\usepackage[noadjust]{cite}

\begin{document}

\title{Developing Optimization-Based Inverter Models for Short Circuit Studies}

\author{\IEEEauthorblockN{Thabiso~R.~Mabote }
\IEEEauthorblockA{\textit{School of Electrical Engineering} \\
\textit{\& Computer Science} \\
\textit{Oregon State University}\\
Corvallis, OR, USA \\
mabotet@oregonstate.edu}
\and
\IEEEauthorblockN{Jose~E.~Tabarez}
\IEEEauthorblockA{\textit{Advanced Network Science Initiative} \\
\textit{Los Alamos National Laboratory}\\
Los Alamos, NM, USA \\
jtabarez@lanl.gov}
\and
\IEEEauthorblockN{Arthur~K.~Barnes}
\IEEEauthorblockA{\textit{Advanced Network Science Initiative} \\
\textit{Los Alamos National Laboratory}\\
Los Alamos, NM, USA \\
abarnes@lanl.gov}
\and
\IEEEauthorblockN{Adam~Mate}
\IEEEauthorblockA{\textit{Advanced Network Science Initiative} \\
\textit{Los Alamos National Laboratory}\\
Los Alamos, NM, USA \\
amate@lanl.gov}
\and
\IEEEauthorblockN{Russell~W.~Bent}
\IEEEauthorblockA{\textit{Advanced Network Science Initiative} \\
\textit{Los Alamos National Laboratory}\\
Los Alamos, NM, USA \\
rbent@lanl.gov}
\and
\IEEEauthorblockN{Eduardo~Cotilla-Sanchez}
\IEEEauthorblockA{\textit{School of Electrical Engineering} \\
\textit{\& Computer Science} \\
\textit{Oregon State University}\\
Corvallis, OR, USA \\
ecs@oregonstate.edu}
}

\IEEEoverridecommandlockouts
\IEEEpubid{\makebox[\columnwidth]{979-8-3503-4743-2/23/\$31.00~\copyright2023 IEEE \hfill} \hspace{\columnsep}\makebox[\columnwidth]{ }}

\maketitle
\IEEEpubidadjcol


\begin{abstract}
As inverter-based generation becomes more common in distribution networks, it is important to create models for use in optimization-based problems that accurately represent their non-linear behavior when saturated.
This work presents models for grid-following and grid-forming inverters, and demonstrates their use in optimization-based fault studies.
The developed models are shown to provide results in line with experimental tests from literature; however, grid-forming inverter models fail to provide feasible solutions for certain fault types even when currents, voltages, and powers from the models seem reasonable. This work lays down the foundation for the development of relaxed models for both grid-following and grid-forming inverter models for use in optimization problems.
\end{abstract}


\section{Introduction} \label{sec:introduction}
\indent

With the increasing adoption of inverter-interfaced generation in microgrids to ensure resilient operation of critical energy infrastructures and meet decarbonization goals, comes the
challenge of ensuring protective coordination. As the size and mixture of distributed energy resources increases in microgrids, so comes the requirement to ensure protection coordination during different modes of operation to minimize the size of affected areas during faults to avoid interruption of critical loads. To achieve this goal, it is important to accurately model the short-circuit current contributions from inverter-interfaced generation for various types of faults and modes of operation \cite{5275777, inverImp}.

Experimental tests have demonstrated that the control schema and pre-fault state of inverters have a profound effect on the fault currents injected by the inverters \cite{8274697, 8669457, 8673877, 9254562, 8980892, 8547488}.
Developing mathematical models to represent the inverters in optimization problems, based on the experimental tests, results in non-linear non-convex constraints leading to feasibility issues \cite{barnes21-pmsp, tabarez22-fccopf-micgrd}. These constraints limit the size and complexity of optimization problems that the inverter models can be used in, especially in optimization-based fault studies used to develop protection coordination.
To overcome this issue, this work investigates the use of Thevenin circuits to approximate continuous mathematical models of grid-forming and grid-following inverters to determine their short-circuit current contributions, and compares the performance of such models to the experimental results found in literature.

The inverter models presented in this work are intended to be used in PowerModelsProtection.jl (PMsP) \cite{barnes21-pmsp}, a structure established on optimization for short-circuit analysis.
PMsP is an extension to a collection of open-source packages using the PowerModels.jl (PMs) \cite{PMs} framework; PMs allows for different optimization problems in power systems to be developed and solved.
Presently, PMsP only produces mixed linear/nonlinear non-convex short-circuit formulations, and it is also dependent on PowerModelsDistribution.jl (PMsD) \cite{PMsD} for creating the optimization-based formulation and mathematical models used for distribution networks.

This work focuses on developing inverter models for use in optimization based short-circuit studies, however, the models are expressed in a general form for use in power flow (pf) and optimal power flow (opf) analysis of balanced and unbalanced distribution networks.
The goal here is to develop continuous variable inverter models that lay the foundation for developing relaxed linear convex approximations in future work.

\vspace{0.05in}
The contributions include the following:
Section~\ref{sec:formulation} presents the optimization-based short-circuit formulation implemented in PMsP.
Section~\ref{sec:inverter-models} presents the developed grid-following and grid-forming inverter models.
Section~\ref{sec:case-study} introduces the case study system used to test and verify the operation of the inverter models, and 
Section~\ref{sec:results} discusses and interprets the observed results.
Last, Section~\ref{sec:conclusions} summarizes the conclusions regarding the state and importance of the work.

\section{Short-Circuit Formulation} \label{sec:formulation}
\normalsize \indent

\textit{The symbols below are used in the following constraint formulation:
\(\mathcal{B}\) is a collection of buses in a system;
\(\mathcal{G}_{ref_b}\) is a collection of reference voltage sources at bus $b$;
\(\Phi_b \subseteq \{A, B, C\}\) is the collection of phases at bus $b$;
\(\mathcal{G}_b\) is a collection of generators at bus $b$;
\(\mathcal{F}\) is the fault in a system;
\(\mathcal{E}_b\) is a collection of lines linked to bus $b$;
\(\mathcal{T}_b\) is a collection of transformers linked to bus $b$; and \(\mathcal{S}_b\) is a collection of shunts at bus $b$.
} \vspace{0.1in}

The optimization-based short-circuit current formulation is implemented using rectangular coordinates inside of PMsP. To indicate the real and imaginary parts the subscripts \(r\) and \(i\) are applied to the variables that are complex. The short-circuit formulation presented here is based on \cite{barnes21-pmsp} and \cite{tabarez22-fccopf-micgrd}.
The formulation defines a set of constraints that are based on Ohm's law and Kirchhoff's current law, which define the phase voltages at the buses due to current flow through various components within the system. These constraints resemble constraints used in power flow, except the load constraints are neglected and constraints that define the faulted condition in the system are added.

\vspace{0.1in}
\normalsize \noindent
Reference Voltage Sources Constraints: \\
\vspace{0.05in}
\small
\noindent $ \forall g \in \mathcal{G}_{ref_n}, ~\forall b \in {\cal B},~\forall \phi \in \Phi_{b}$
\begin{subequations}
\label{const:ref_volt_scr}
\begin{align}
&V^\phi_{g_r} = V^\phi_{g_{setp.}} \cdot cos \left( \theta_{g_{setp.}} \right) & \\
&V^\phi_{g_i} = V^\phi_{g_{setp.}} \cdot sin \left( \theta_{g_{setp.}} \right) & \\
&V^\phi_{b_r} = V^\phi_{g_r} - r^\phi_g \cdot I^\phi_{g_r} + x^\phi_g \cdot I^\phi_{g_i}& \\
&V^\phi_{b_i} = V^\phi_{g_i} - r^\phi_g \cdot I^\phi_{g_i} - x^\phi_g \cdot I^\phi_{g_r}&
\end{align}
\end{subequations}
\vspace{0.05in}

\normalsize \noindent
Generator Power Constraints: \\
\vspace{0.05in}
\small
\noindent $\forall g \in {\cal G},~\forall b \in {\cal B},~\forall \phi \in \Phi_{b}$
\begin{subequations}
\label{const:constant_gen}
\begin{align}
V^\phi_{g_r} \cdot I^\phi_{g_r} + V^\phi_{g_i} \cdot I^\phi_{g_i} &= P^\phi_{g} \\
V^\phi_{g_i} \cdot I^\phi_{g_r} - V^\phi_{g_r} \cdot I^\phi_{g_i} &= Q^\phi_{g}\\
\Delta P^\phi_{g} &\le 5\% \\
\Delta Q^\phi_{g} &\le 5\% \\
-\frac{P^\phi_{g}}{pf} \cdot \sin(\textrm{acos}(pf)) \le Q^\phi_{g} &\le \frac{P^\phi_{g}}{pf} \cdot \sin(\textrm{acos}(pf))
\end{align}
\end{subequations}
\vspace{0.05in}

\normalsize \noindent
Fault Current Constraints: \\
\vspace{0.05in}
\small
\noindent $ \forall f_b \in \mathcal{F}, ~\forall b \in {\cal B},~\forall \phi \in \Phi_{b}$
\begin{subequations}
\label{const:fault_current}
\begin{align}
I^\phi_{{f_b}_r} = \sum_{c \in \Phi} \left( G_{\phi,c} \cdot V^c_{b_r} \right) \\
I^\phi_{{f_b}_i} = \sum_{c \in \Phi} \left( G_{\phi,c} \cdot V^c_{b_i} \right)
\end{align}
\end{subequations}
\vspace{0.05in}

\normalsize \noindent
Kirchhoff's Current Constraints to Unfaulted Buses: \\
\vspace{0.05in}
\small
\noindent $\forall b \in {\cal B},~\forall \phi \in \Phi_{b}$
\begin{subequations}
\label{const:kirch_i_nonf}
\begin{align}
&\sum_{(b,j) \in \mathcal{E}_b} I^\phi_{{(b,j)}_r} + \sum_{(b,j) \in \mathcal{T}_b} I^\phi_{{(b,j)}_r}= \sum_{g \in \mathcal{G}_b} I^\phi_{{g}_r} + \sum_{s \in \mathcal{S}_b} I^\phi_{{s}_r} &\\
&\sum_{(b,j) \in \mathcal{E}_b}I^\phi_{{(b,j)}_i} + \sum_{(b,j) \in \mathcal{T}_b} I^\phi_{{(b,j)}_i}= \sum_{g \in \mathcal{G}_b} I^\phi_{{g}_i} + \sum_{s \in \mathcal{S}_b} I^\phi_{{s}_i} & 
\end{align}
\end{subequations}
\vspace{0.05in}

\normalsize \noindent
Kirchhoff's Current Constraints to Faulted Buses: \\
\vspace{0.05in}
\noindent $\forall b \in {\cal B},~\forall \phi \in \Phi_{b}$
\begin{subequations}\label{const:kirch_i_f}
\begin{align}
&\sum_{b,j \in \mathcal{E}_b} I^\phi_{{(b,j)}_r} + \sum_{b,j \in \mathcal{T}_b} I^\phi_{{(b,j)}_r}= \sum_{g \in \mathcal{G}_b} I^\phi_{{g}_r} + \sum_{s \in \mathcal{S}_b} I^\phi_{{s}_r} + I^\phi_{{f}_r}&\\
&\sum_{b,j \in \mathcal{E}_b} I^\phi_{{(b,j)}_i} + \sum_{b,j \in \mathcal{T}_b} I^\phi_{{(b,j)}_i}= \sum_{g \in \mathcal{G}_b} I^\phi_{{g}_i} + \sum_{s \in \mathcal{S}_b} I^\phi_{{s}_i} + I^\phi_{{f}_i}& 
\end{align}
\end{subequations}
\vspace{0.05in}

\normalsize \noindent
Voltage Drop Constraints: \\
\vspace{0.05in}
\small
\noindent $\forall (b,j) \in {\cal E},~\forall b \in {\cal B},~\forall \phi \in \Phi_{b}$
\begin{subequations}
\label{const:volt_drop}
\begin{align}
&V^\phi_{{b}_r} = V^\phi_{{b}_r} - r^\phi_{(b,j)} \cdot I^\phi_{{(b,j)}_r} + x^\phi_{(b,j)} \cdot I^\phi_{{(b,j)}_i}& \\
&V^\phi_{{b}_i} = V^\phi_{{b}_i} - r^\phi_{(b,j)} \cdot I^\phi_{{(b,j)}_i} - x^\phi_{(b,j)} \cdot I^\phi_{{(b,j)}_r}& \\
&V^\phi_{b_{min}} \le (V^\phi_{b_r})^2 + (V^\phi_{b_i})^2 \le V^\phi_{b_{max}} & \\
& (I^\phi_{{(b,j)}_r})^2 + (I^\phi_{{(b,j)}_i})^2 \le I^\phi_{(b,j)_{thermal}} &
\end{align}
\end{subequations}
\vspace{0.05in}

\normalsize \noindent
Transformer Constraints: \\
\vspace{0.05in}
\small
\noindent $\forall (b,j) \in {\cal T},~\forall b \in {\cal B},~\forall \phi \in \Phi_{b}$
\begin{subequations}
\label{const:ofp_transformer_thermal}
\begin{align}
&W_{(b,j)} \cdot V^\phi_{{b}_r} = \eta_{(b,j)} \cdot V^\phi_{{j}_r} & \\
&W_{(b,j)} \cdot V^\phi_{{b}_i} = \eta_{(b,j)} \cdot V^\phi_{{j}_i} & \\
&I^\phi_{{b}_i} = \eta_{(b,j)} I^\phi_{{j}_i} & \\
&(I^\phi_{{(b,j)}_r})^2 + (I^\phi_{{(b,j)}_i})^2 \le I^\phi_{(b,j)_{thermal}} &
\end{align}
\end{subequations}
\vspace{0.05in}

\normalsize
The fault currents, as shown in Constraints~(\ref{const:fault_current}a)-(\ref{const:fault_current}b), are derived using a fault admittance matrix, which denotes a fault impedance between phase and ground.
Line-to-line and 3-phase to ground faults cannot be modeled directly due to the presence of the impedances between the phases and ground, consequently an equivalent impedance admittance matrix is derived using a star-mesh transformation \cite{barnes21-pmsp}. Along with fault admittance constraints and ~Constraints~(\ref{const:kirch_i_nonf}) and (\ref{const:kirch_i_f}), which couples the nodes of the network using Kirchhoff's current law, allows for the formulation not to require an explicit objective function.
Constraints~(\ref{const:ref_volt_scr}a)-(\ref{const:ref_volt_scr}d) represents the reference voltage sources; normally the distribution substation is the main voltage source in grid-connected mode of operation, whereas it is modeled as a voltage source behind an impedance during a fault, and this impedance is obtained from the single-phase and 3-phase short-circuit powers (Constraint~(\ref{const:ref_volt_scr}b)).
Synchronous generation within the network is defined by Constraints~(\ref{const:constant_gen}a)-(\ref{const:constant_gen}e). 
The last set of constraints, Constraints~(\ref{const:ofp_transformer_thermal}a)-(\ref{const:ofp_transformer_thermal}d), define the voltage and current changes across lines and transformers within the network.

\section{Inverter Models} \label{sec:inverter-models}
\normalsize \indent

\begin{figure}[!htbp]
\centering
\includegraphics[width=0.35\textwidth]{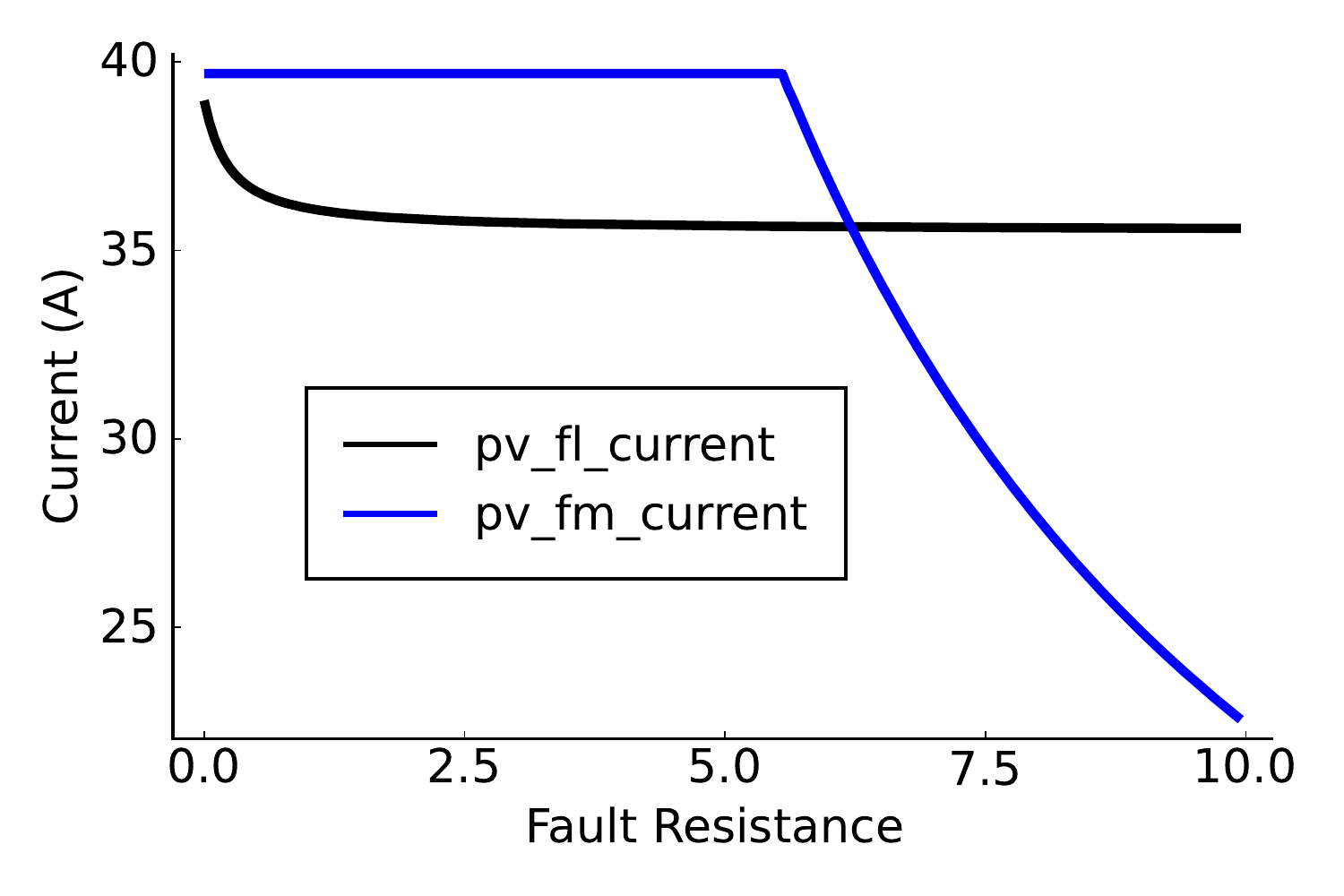}
\caption{Fault Current (A) against Fault Resistance (R) for pv\_fl\_current (grid-following) model and pv\_fm\_current (grid-forming) model.}
\label{fig:pmps-init-lowfaultimpedance}
\end{figure}

\begin{figure}[!htbp]
\centering
\includegraphics[width=0.35\textwidth]{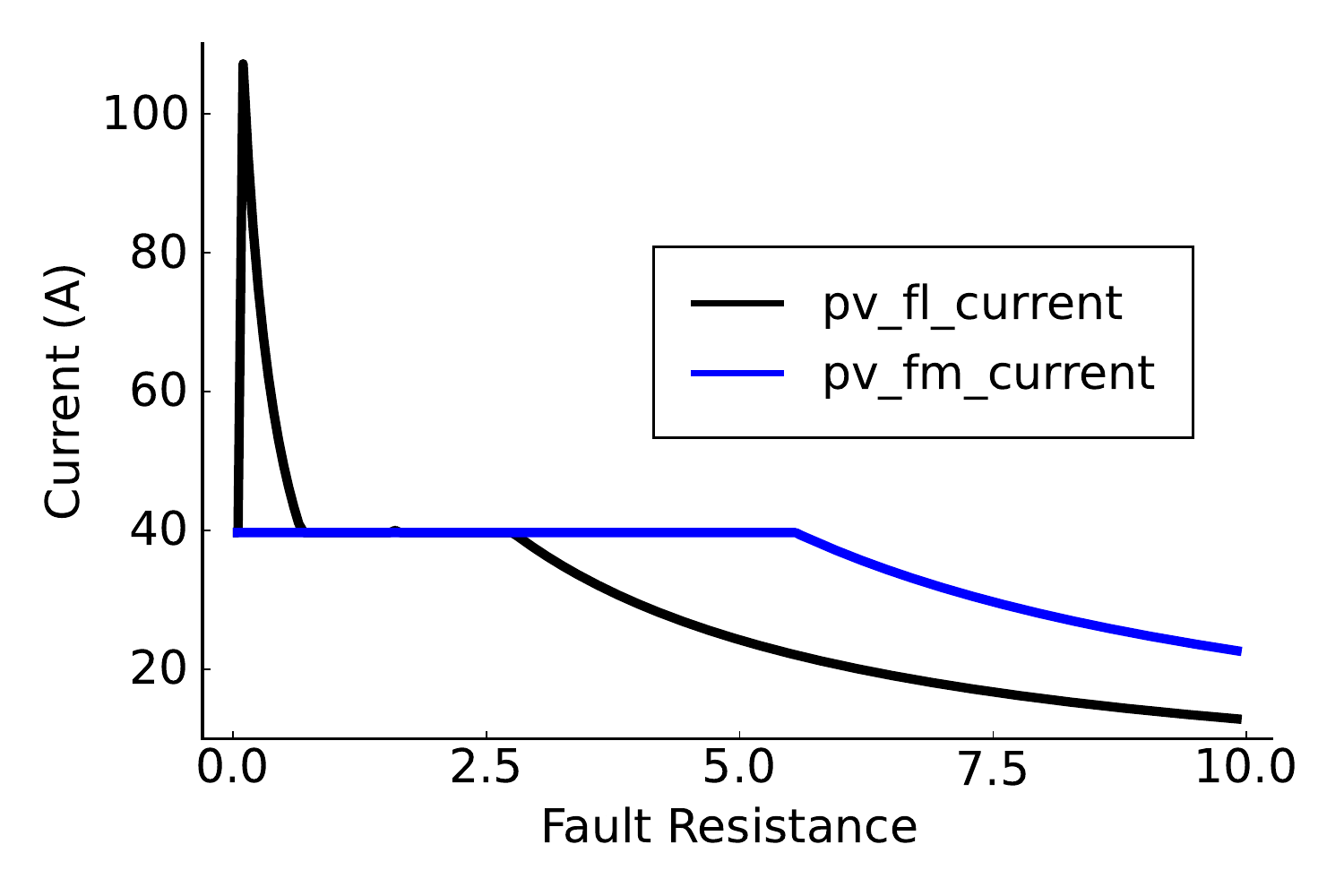}
\caption{Fault Current (A) against Fault Resistance (R) for pv\_fl\_current (grid-following) model and pv\_fm\_current (grid-forming) model.}
\label{fig:pmps-init-closerfault}
\end{figure}

The models presented in \cite{barnes21-pmsp} define inverters with complex mathematical models that result in infeasible solutions under certain fault conditions.
This can be observed in Fig.~\ref{fig:pmps-init-lowfaultimpedance} and Fig.~\ref{fig:pmps-init-closerfault}: for low fault impedances located away from the inverter (Fig.~\ref{fig:pmps-init-lowfaultimpedance}), the optimization solver is able to find a feasible solution for both grid-forming and grid-following inverters; however, as the fault was moved closer to the inverters (Fig.~\ref{fig:pmps-init-closerfault}), the grid-following inverter model resulted in infeasible solutions causing large deviations in current.

This work aims to develop simplified inverter models based on Thevenin approximations for more general use in fault studies.
Thevenin approximations are widely used to model generation sources in fault studies; the difficulty is to incorporate the inverter's non-linear current response within them.
When inverters are not saturated, they respond similarly to the Thevenin models. Incorporating the current saturation into the mathematical constraints of the optimization problem is difficult when trying to use continuous decision variables.

The developed inverter models in this work are compared to experimental results to evaluate their performance on different fault types.
It must be noted that the models presented below do not directly model the grid interfacing transformers; the constraints for the inverters are defined such that they help compensate for the transformer models not being directly modeled.
\vspace{1cm}

\subsection{Grid-Following Inverter Model}
\normalsize \indent
\vspace{.3cm}

The grid-following inverter model was designed to act as a constant-power source up to the point where the inverter's current saturates, then become a constant-current source on the saturated phase.

\vspace{0.05in}
\small
\noindent $\forall g \in {\cal G}_{gfli},~\forall b \in {\cal B},~\forall \phi \in \Phi_{b}$
\begin{subequations}
\label{const:constant_gfli}
\begin{align}
&{I^\phi_{g_r}}^2 +  {I^\phi_{g_i}}^2 \le {I^\phi_{g_{max}}}^2 & \\
&V^\phi_{g_r} \cdot I^\phi_{g_r} + V^\phi_{g_i} \cdot I^\phi_{g_i} = P^\phi_{g} - z^\phi & \\
&M\cdot (1-\frac{1}{{I^\phi_{g_{max}}}^2} ({I^\phi_{g_r}}^2 +  {I^\phi_{g_i}}^2)) \cdot z^\phi \ge 0 & \\
&V^\phi_{g_r} \cdot I^\phi_{g_r} + V^\phi_{g_i} \cdot I^\phi_{g_i} \ge 0 & \\
&V^\phi_{g_i} \cdot I^\phi_{g_r} - V^\phi_{g_r} \cdot I^\phi_{g_i} \le Q^\phi_{g_{max}} & \\
&V^\phi_{g_i} \cdot I^\phi_{g_r} - V^\phi_{g_r} \cdot I^\phi_{g_i} \ge Q^\phi_{g_{min}} & \\
&0 \le z^\phi \le 1 &
\end{align}
\end{subequations}
\vspace{0.05in}

\normalsize
Constraint~(\ref{const:constant_gfli}a) ensures that the inverter-sourced current is below the maximum possible sourced current \(I^\phi_{g_{max}}\) that the inverter can supply, which is a parameter defined by the User or set by PMsP.
Constraint~(\ref{const:constant_gfli}b) defines the active power sourced from the inverter as a function of bus voltage and inverter current; \(P^\phi_{g}\) is the set-point for the active power per phase of the inverter.
Under non-faulted or light faulted cases the inverter will supply the defined power, but as a fault drives down the voltage, the current will increase to meet the power requirement; once the current reaches its limit, Constraint~(\ref{const:constant_gfli}c) activates the continuous variable Z, which in Constraint~(\ref{const:constant_gfli}b) allows the inverter to drop its active power output based on the voltage collapse due to a fault. 
The last set of constraints, Constraints~(\ref{const:constant_gfli}d)-(\ref{const:constant_gfli}f), ensure that active power is supplied by the inverter and never absorbed, and that the inverter operates within a fixed power factor range by setting the limits on the reactive power supplied or absorbed.

\subsection{Grid-Forming Inverter Model}
\normalsize \indent

Two Thevenin approximation models were developed and tested to simulate a grid-forming inverter.
The first is a simplified model of a voltage source behind an adjustable impedance.
The second is a voltage source that can adjust its voltage magnitude on each phase.
\vspace{.6cm}

\vspace{0.1in} \noindent
\textit{Simplified Grid-Forming Inverter}
\vspace{0.1in} \indent

\vspace{0.05in}
\small
\noindent $\forall g \in {\cal G}_{gfmi},~\forall b \in {\cal B},~\forall \phi \in \Phi_{b}$
\begin{subequations}
\label{const:constant_gfmi_simp}
\begin{align}
&{I^\phi_{g_r}}^2 + {I^\phi_{g_i}}^2 \le {I^\phi_{g_{max}}}^2 & \\
&M\cdot (1-\frac{1}{{I^\phi_{g_{max}}}^2} ({I^\phi_{g_r}}^2 +  {I^\phi_{g_i}}^2)) \cdot r^\phi \ge 0 & \\
&V^\phi_{g_r}= V^\phi_{g_{r0}} - r^\phi \cdot I^\phi_{g_r}& \\
&V^\phi_{g_i}= V^\phi_{g_{i0}} - r^\phi \cdot I^\phi_{g_i}& \\
&0 \le  r^\phi \le \frac{1}{I^\phi_{g_{max}}}& 
\end{align}
\end{subequations}
\vspace{0.05in}

\normalsize
Constraint~(\ref{const:constant_gfmi_simp}a) ensures that the inverter sourced current is below the maximum possible sourced current \(I^\phi_{g_{max}}\).
Constraint~(\ref{const:constant_gfmi_simp}b) makes the resistor an activation variable that is zero when the current is below \(I^\phi_{g_{max}}\) and is non-zero when the inverter is saturated in order to allow the inverter's bus voltage to collapse.
Constraints~(\ref{const:constant_gfmi_simp}c) and (\ref{const:constant_gfmi_simp}d) define the Thevenin approximation, using a resistor as the Thevenin impedance; the value of the resistor is determined by the solver as a value between 0 and a maximum value based on the maximum current supplied by the inverter (Constraint~(\ref{const:constant_gfmi_simp}e)).
\vspace{.6cm}

\vspace{0.1in} \noindent
\textit{Complex Grid-Forming Inverter}
\vspace{0.1in} \indent
\vspace{.3cm}

\normalsize
Constraint~(\ref{const:constant_gfmi}a) ensures that the inverter sourced current is below the maximum possible sourced current \(I^\phi_{g_{max}}\).
Constraints~(\ref{const:constant_gfmi}c) and (\ref{const:constant_gfmi}d) define the inverter voltage applied to a bus:~\(V^\phi_{o_g}\) defines the voltage magnitude of the inverter pre-fault, which has to be equal to the bus voltage when the current is not saturated; when the current saturates, the activation variable \(z^\phi\) allows the inverter's voltage magnitude to adjust (Constraint~(\ref{const:constant_gfmi}b).
Constraints~(\ref{const:constant_gfmi}e)-(\ref{const:constant_gfmi}g) enforce balanced phase angles across the inverter's voltages.
The last set of constraints, Constraints~(\ref{const:constant_gfmi}h)-(\ref{const:constant_gfmi}j), ensure that the inverter is operating within its power limits.

\vspace{0.05in}
\small
\noindent $\forall g \in {\cal G}_{gfmi},~\forall b \in {\cal B},~\forall \phi \in \Phi_{b}$
\begin{subequations}
\label{const:constant_gfmi}
\begin{align}
&{I^\phi_{g_r}}^2 + {I^\phi_{g_i}}^2 \le {I^\phi_{g_{max}}}^2 & \\
&M\cdot (1-\frac{1}{{I^\phi_{g_{max}}}^2} ({I^\phi_{g_r}}^2 +  {I^\phi_{g_i}}^2)) \cdot z^\phi \ge 0 & \\
&{V^\phi_{g_r}}^2 + {V^\phi_{g_i}}^2 \le {V^\phi_{g_0}}^2 \cdot (1+ z^\phi)& \\
&{V^\phi_{g_r}}^2 + {V^\phi_{g_i}}^2 \ge {V^\phi_{g_0}}^2 \cdot (1- z^\phi)& \\
&V^\phi_{g_r} \cdot V^\phi_{g_i0} - V^\phi_{g_r} \cdot V^\phi_{g_z^\phi} = 0& \\
&V^\phi_{g_r} \cdot V^\phi_{g_r0} \ge 0 & \\
&V^\phi_{g_i} \cdot V^\phi_{g_i0} \ge 0 & \\
&\sum_{\phi \in \Phi} \left ( V^\phi_{g_r} \cdot I^\phi_{g_r} + V^\phi_{g_i} \cdot I^\phi_{g_i} \right )= P& \\
&\sum_{\phi \in \Phi} \left ( V^\phi_{g_i} \cdot I^\phi_{g_r} - V^\phi_{g_r} \cdot I^\phi_{g_i} \right )= Q& \\
&P + Q \le S_{max} &\\
&0 \le z^\phi \le 1 &
\end{align}
\end{subequations}
\vspace{0.05in}

\section{Case Study System} \label{sec:case-study}
\normalsize \indent

\begin{figure}[!htbp]
\centering
\includegraphics[width=0.4\textwidth]{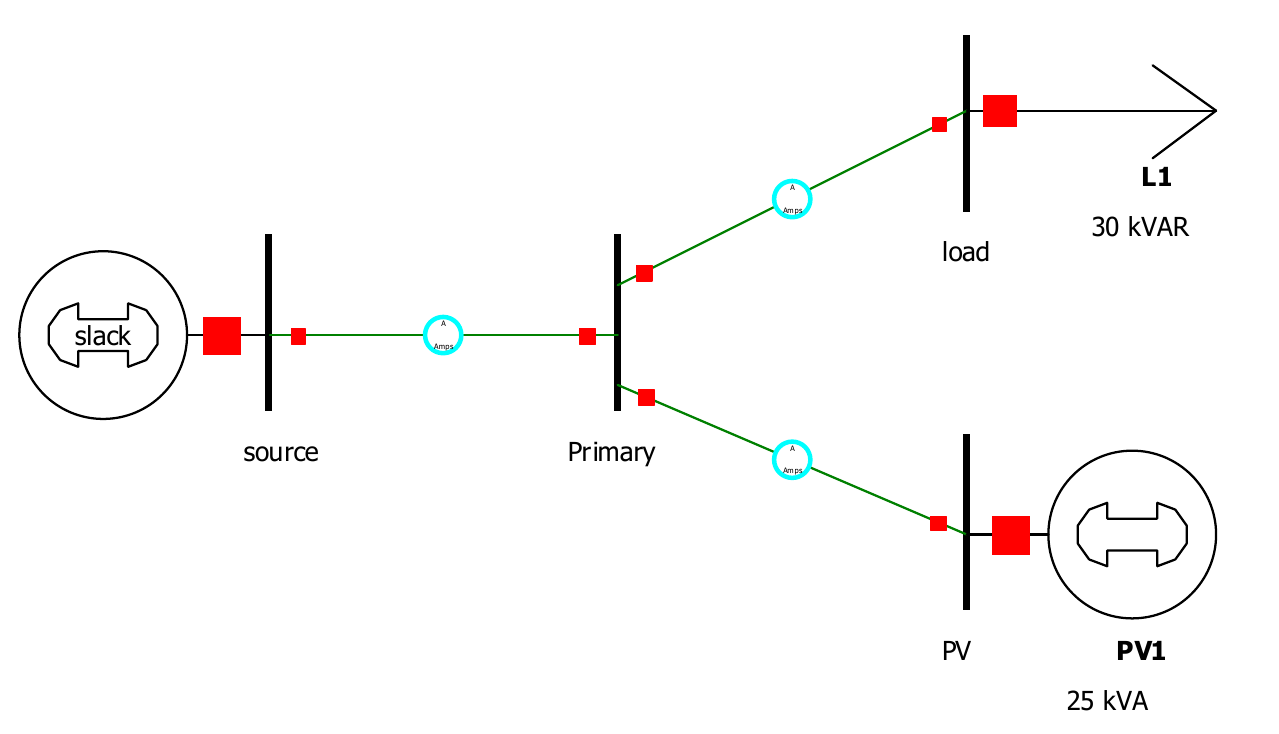}
\caption{One-line diagram of the 4-bus case study system.}
\label{fig:case33}
\end{figure}

The case study system used to analyze and verify the operation of the developed inverter models is a 4-bus synthetic microgrid, with photovoltaic solar inverter-interfaced distributed generation.
Fig.~\ref{fig:case33} presents the one-line diagram of the system; it can be accessed in the PMsP GitHub as \texttt{case3\_balanced\_pv.dss}\cite{noauthor_powermodelsprotectionjltestdatadist_nodate}.

The grid-following and grid-forming inverters were tested independently.
Each was installed on the \textit{PV~Bus}, and line-to-line, line-to-ground, and 3-phase-to-ground faults were applied to the \textit{Load Bus}, which is electrically close to the inverters.
The grid-following inverter was set to have a rated power of 25 kVA, while the grid-forming inverter to 100 kVA; the reason for the difference is to ensure that the grid-forming inverter's current was not constantly saturated, trying to energize the circuit and feed the fault, since in the tests the line between the substation (\textit{Source Bus}) and the \textit{Primary Bus} was opened to create an island.

\section{Results and Discussion} \label{sec:results}
\normalsize \indent

The fault results of the developed inverter models were compared to the experimental results presented in \cite{8274697, 8669457, 8673877, 9254562, 8980892, 8547488}.
To evaluate these new inverter models, sweeps of fault impedance were performed to produce plots of fault currents, voltages, and active and reactive powers versus the fault impedances.

\subsection{Grid-Following Inverter Model}
\normalsize \indent

\begin{figure}[!htbp]
    \subfigure{
    \includegraphics[width=0.46\linewidth]{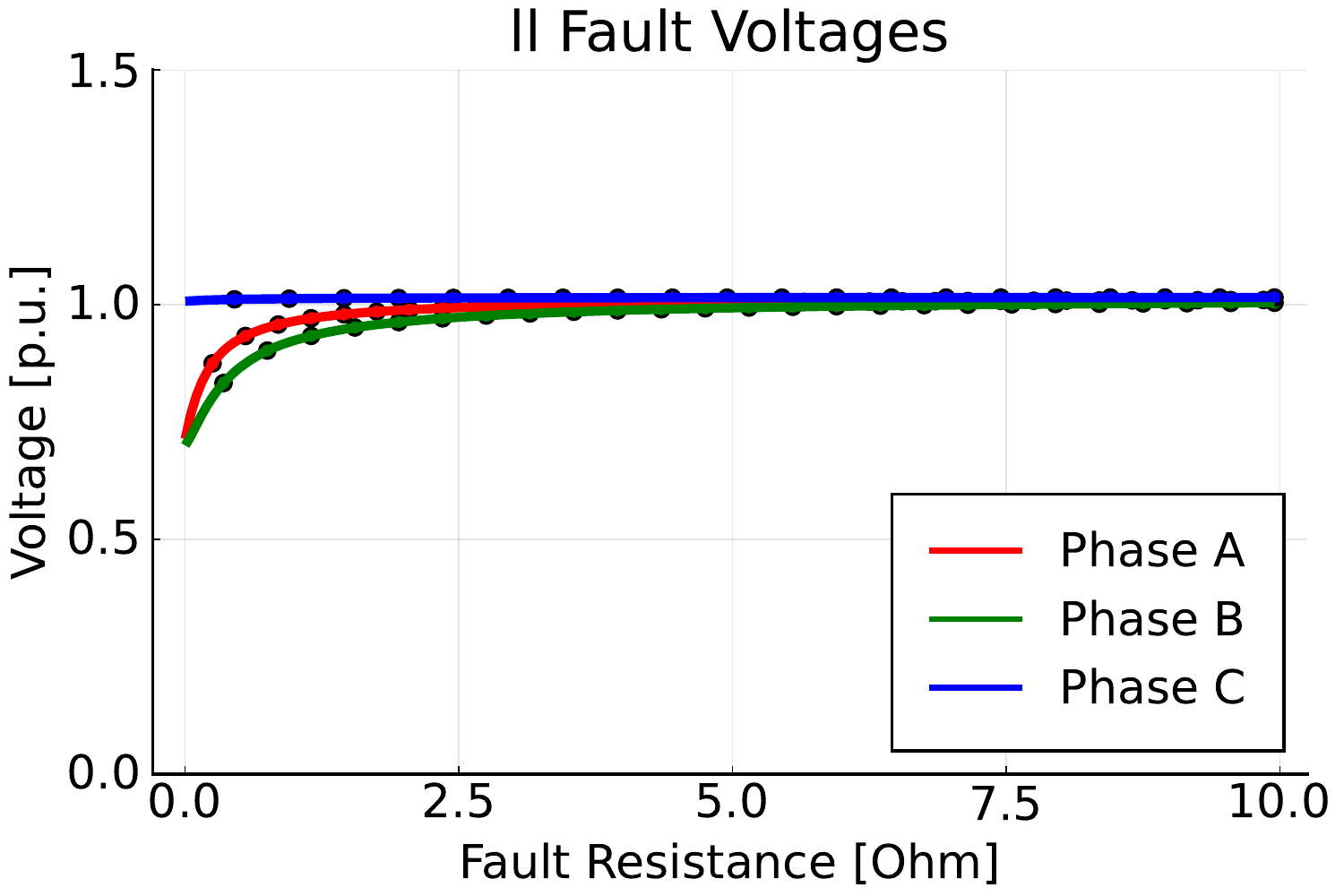}
    }
    \subfigure{
    \includegraphics[width=0.46\linewidth]{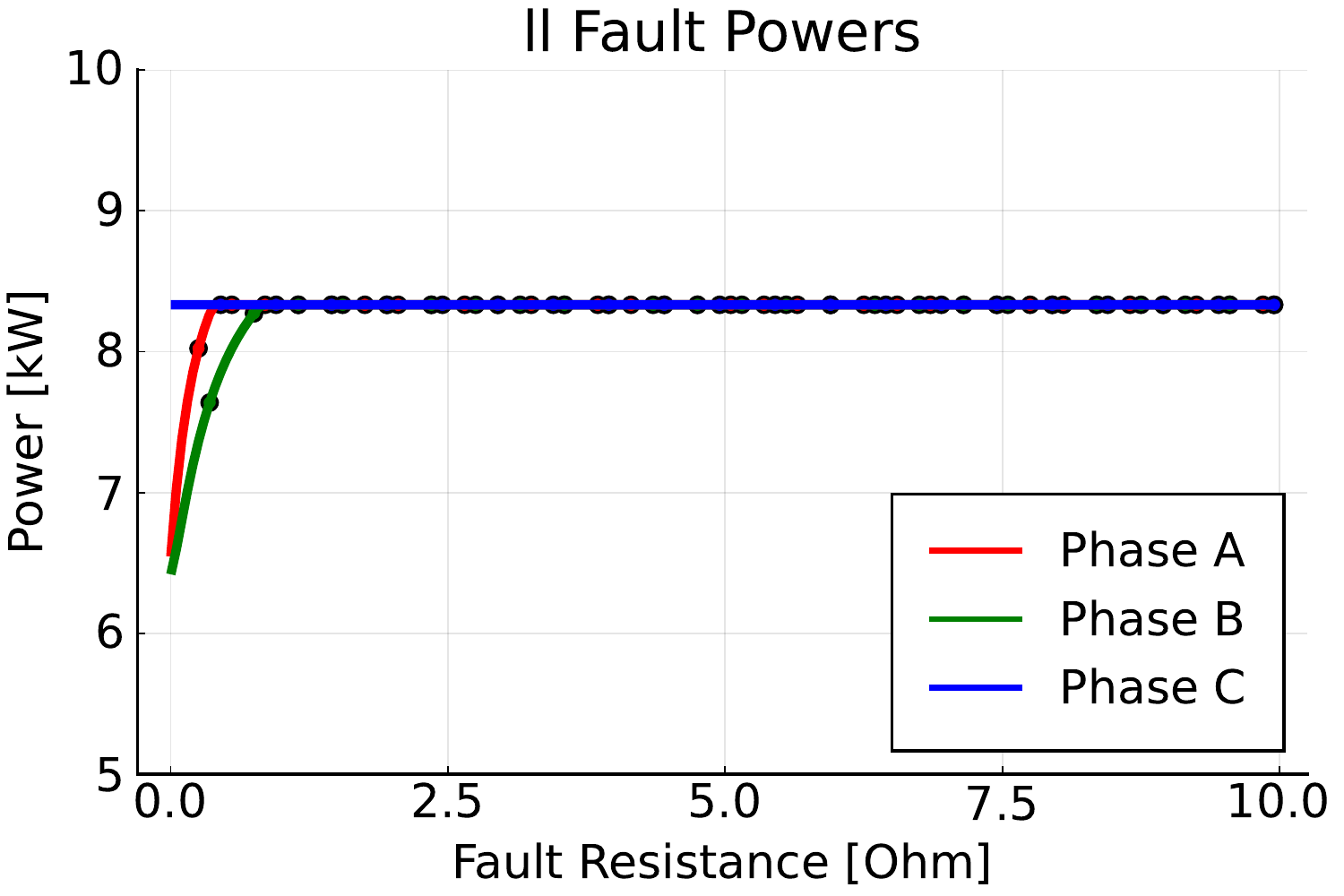}
    }
    \subfigure{
    \includegraphics[width=0.46\linewidth]{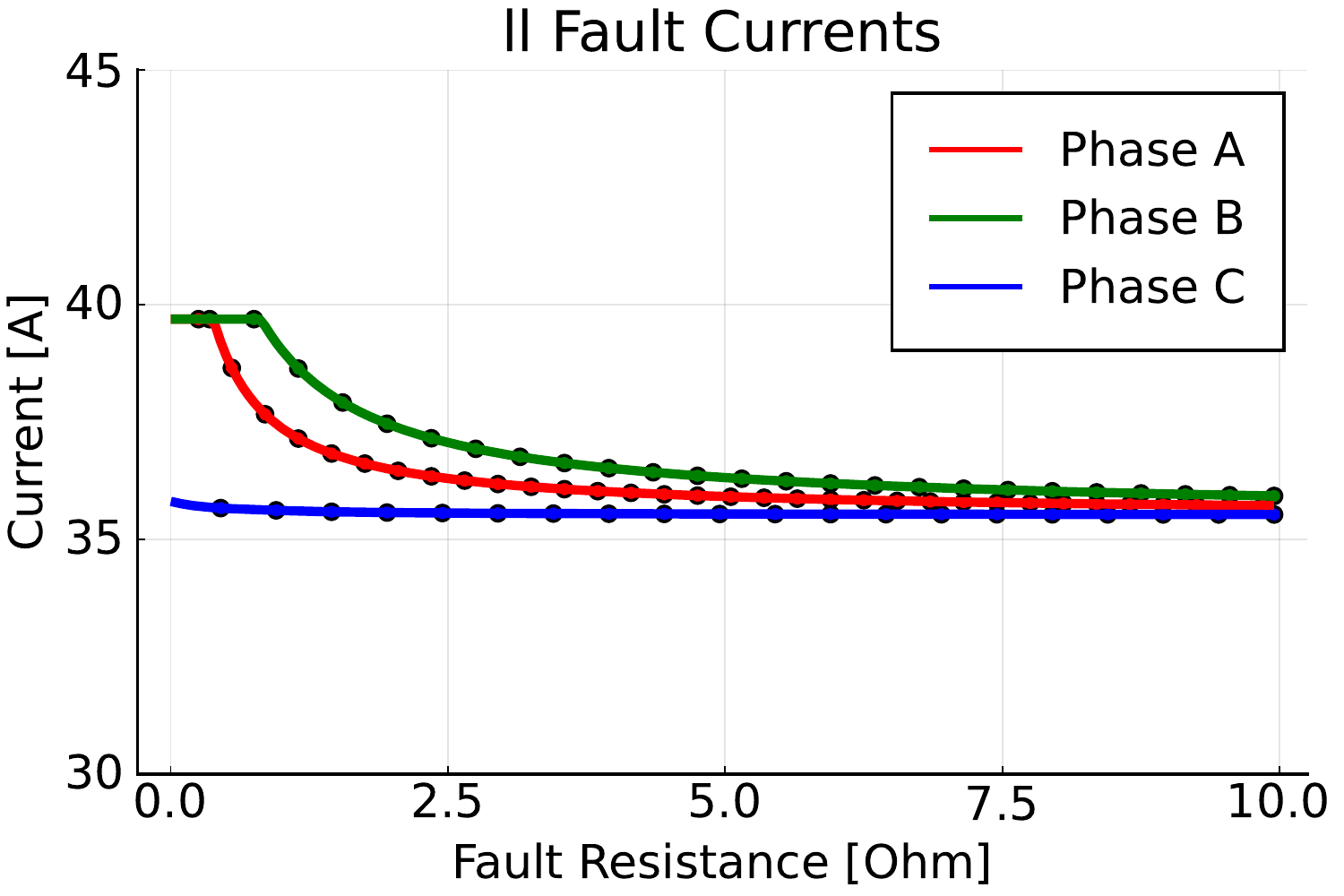}
    }
     \subfigure{
    \includegraphics[width=0.46\linewidth]{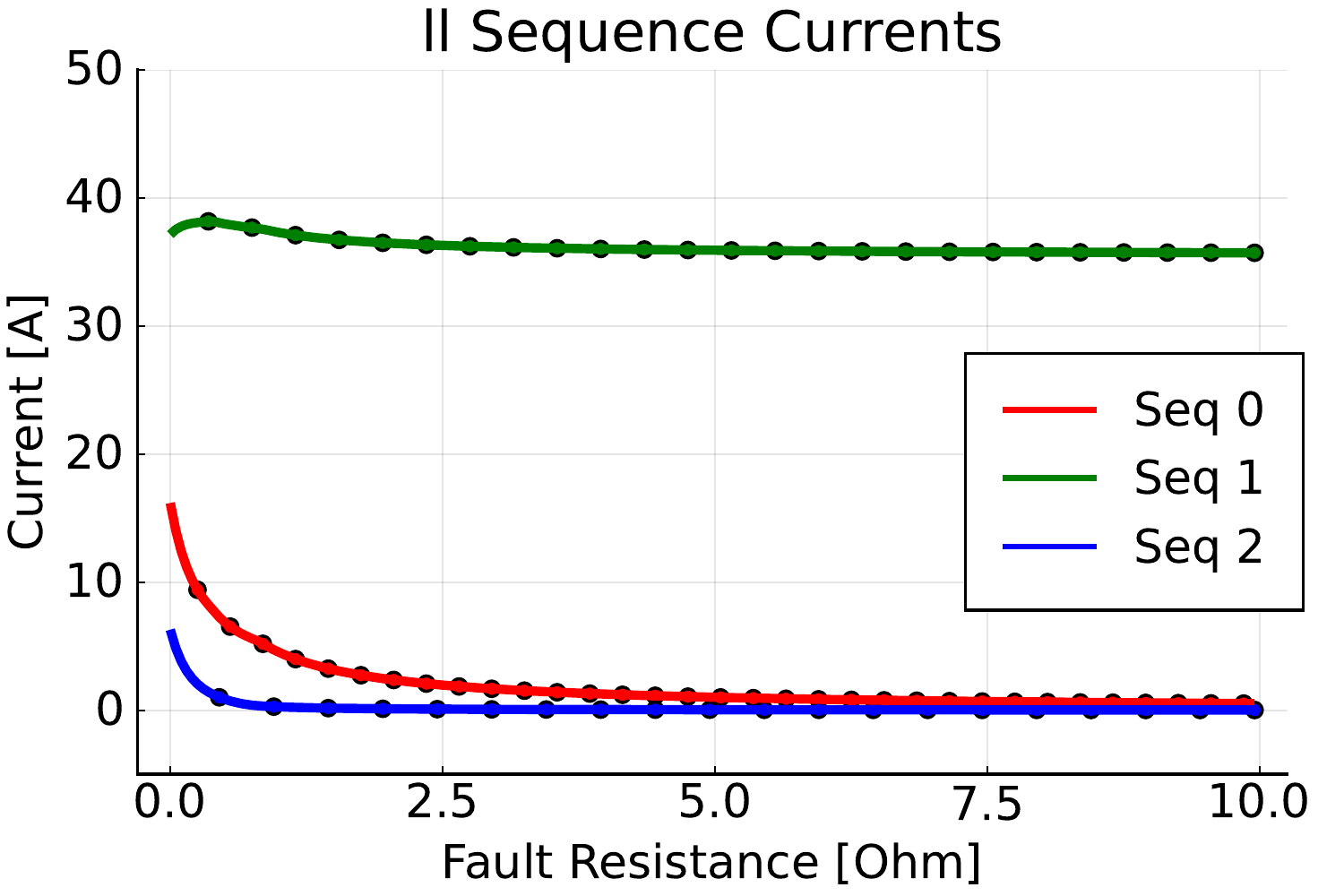}
    }
    \caption{Grid-Following Inverter -- Fault Currents [$A$] For Line-Line Fault.}
    \label{fig:grid-follow-line-line}
\end{figure}

\begin{figure}[!htbp]
    \subfigure{
    \includegraphics[width=0.46\linewidth]{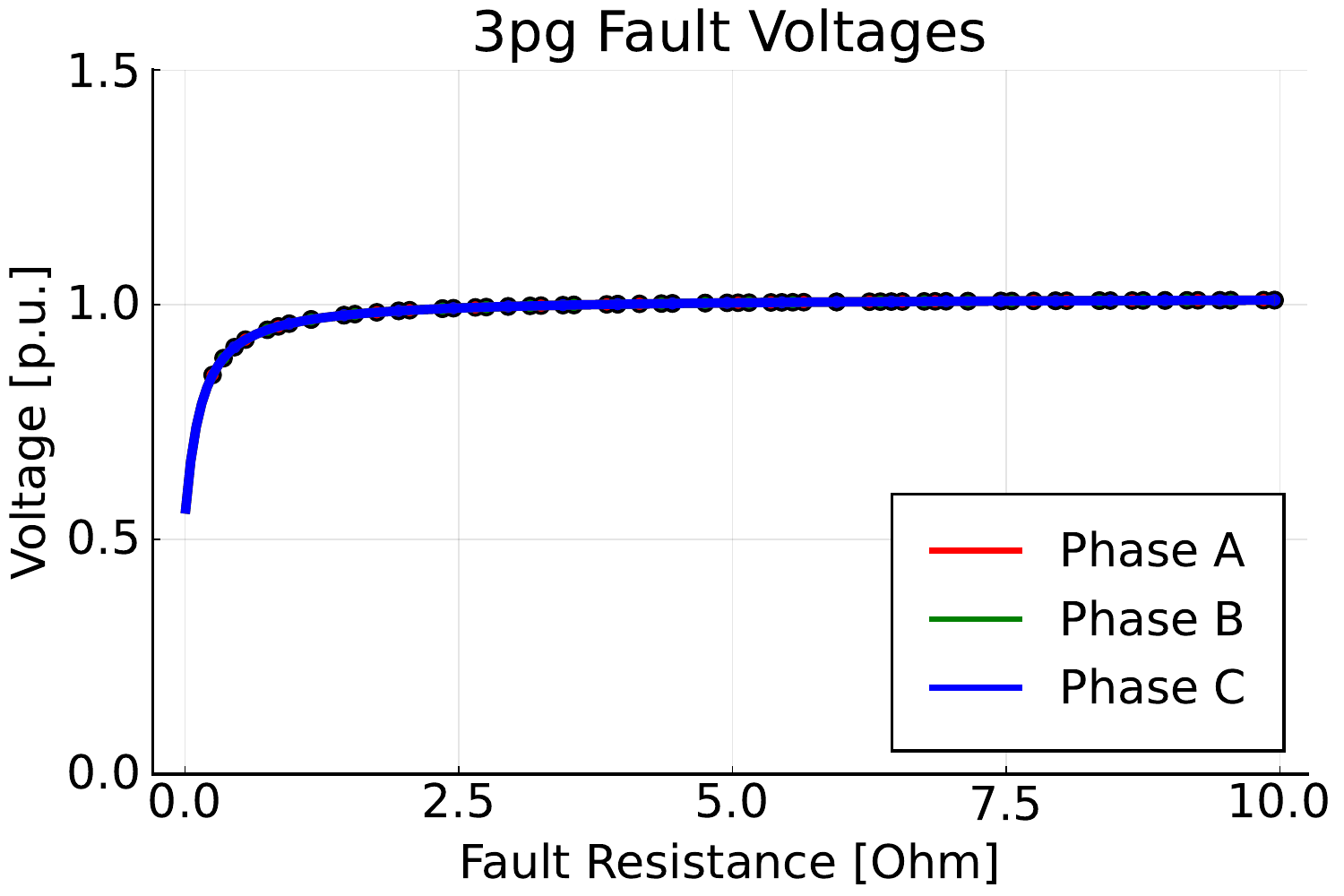}
    }
    \subfigure{
    \includegraphics[width=0.46\linewidth]{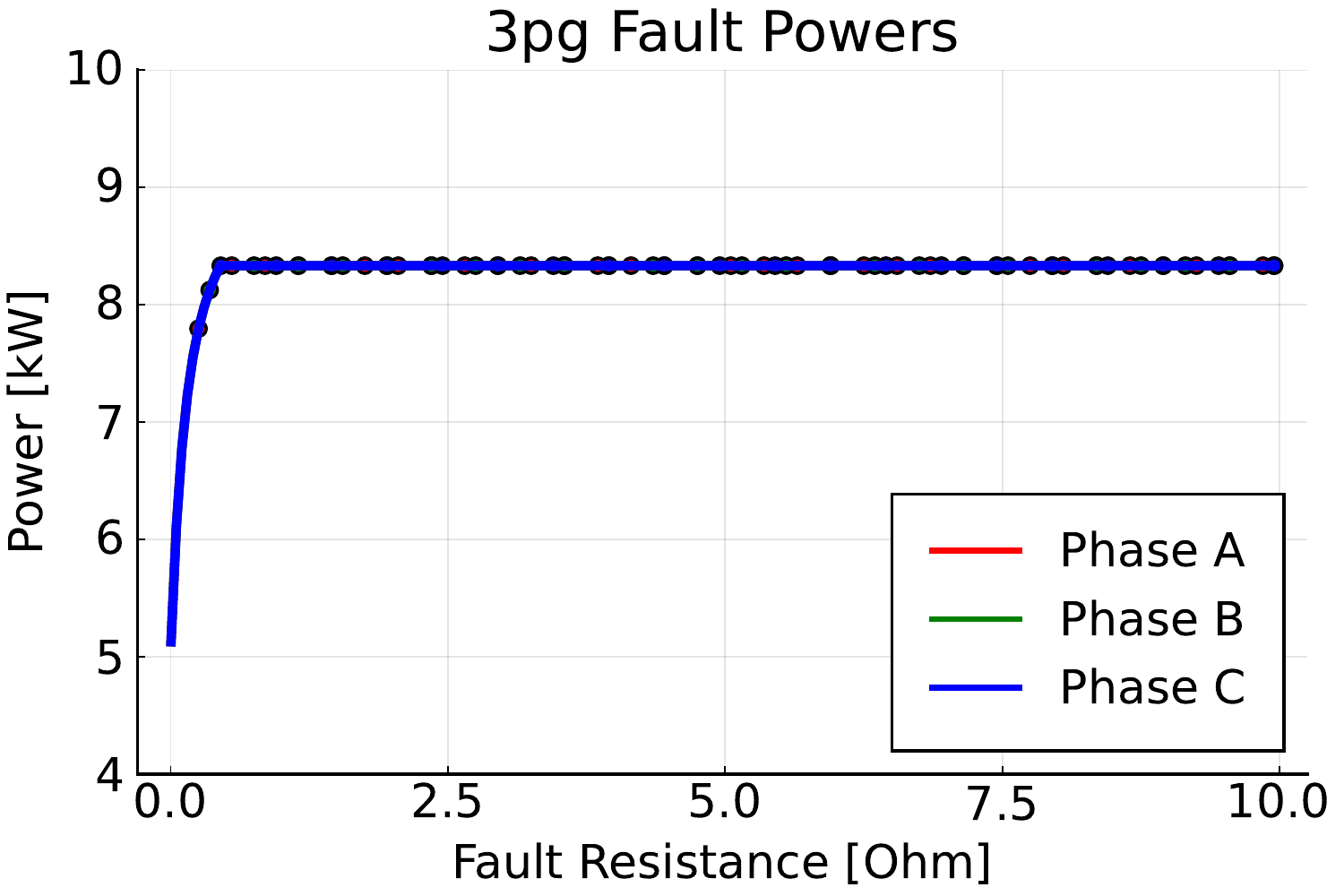}
    }
    \subfigure{
    \includegraphics[width=0.46\linewidth]{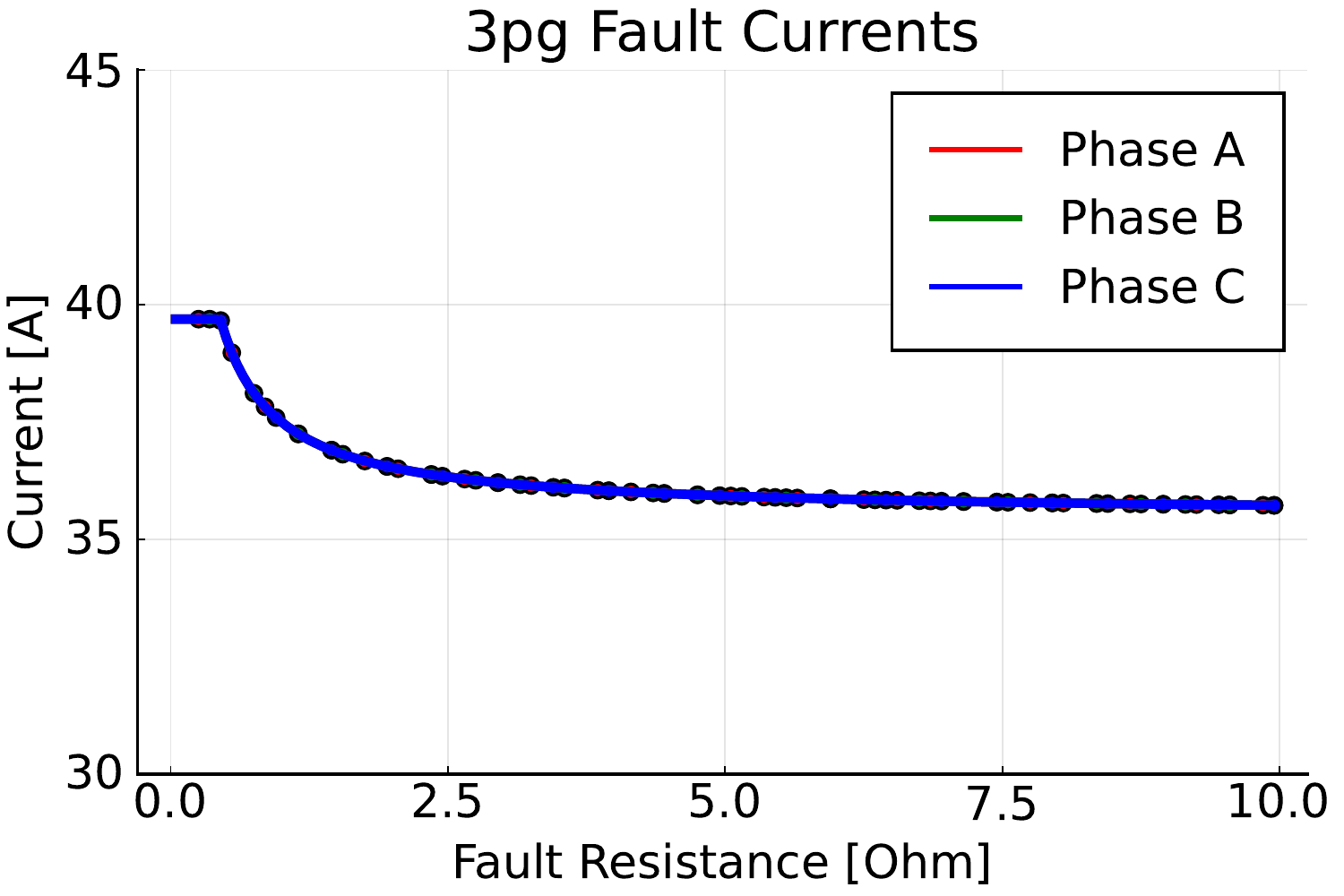}
    }
     \subfigure{
    \includegraphics[width=0.46\linewidth]{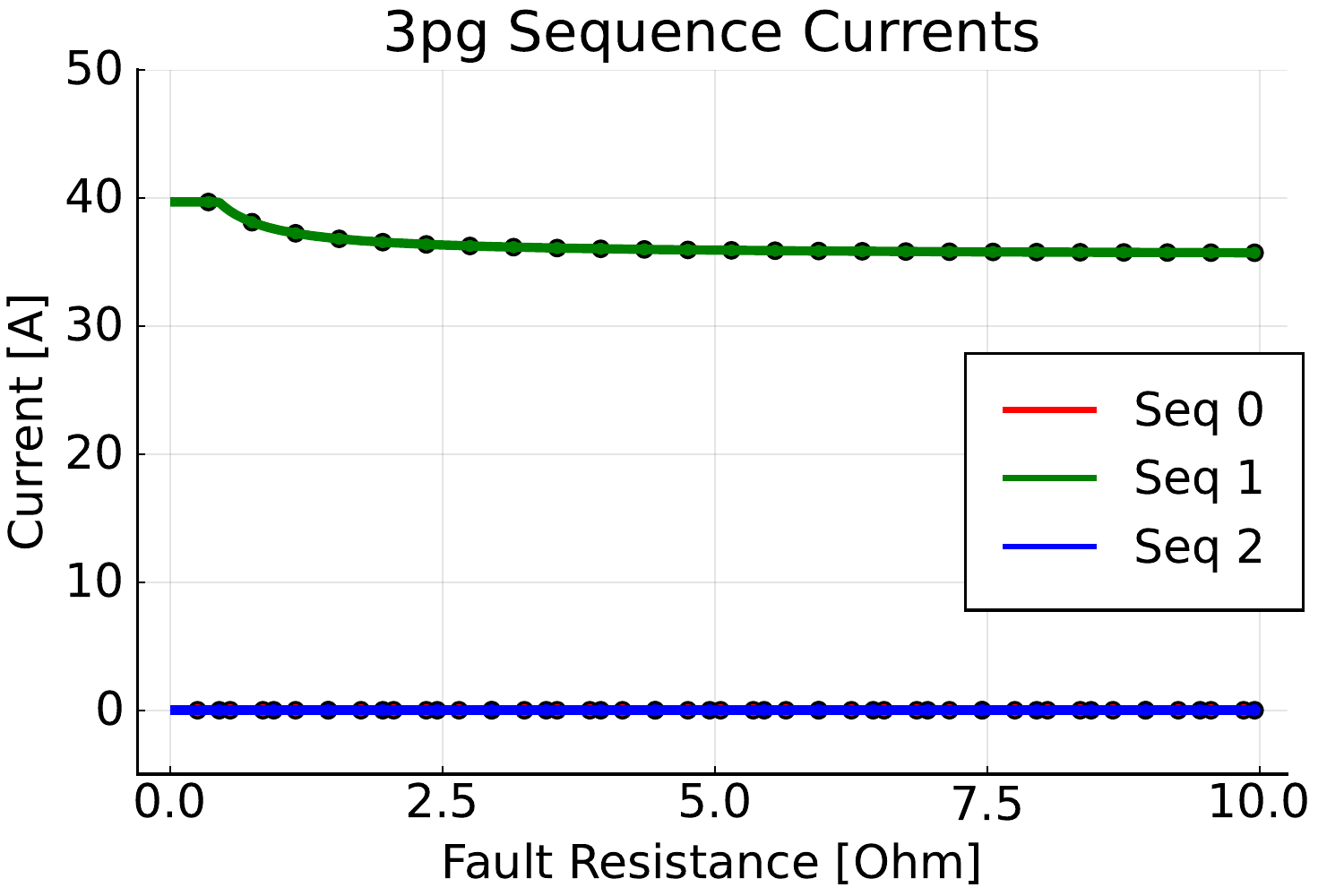}
    }
    \caption{Grid-Following Inverter -- Fault Currents [$A$] For 3-Phase Fault.}
    \label{fig:grid-follow-3ph}
\end{figure}

\begin{figure}[!htbp]
    \subfigure{
    \includegraphics[width=0.46\linewidth]{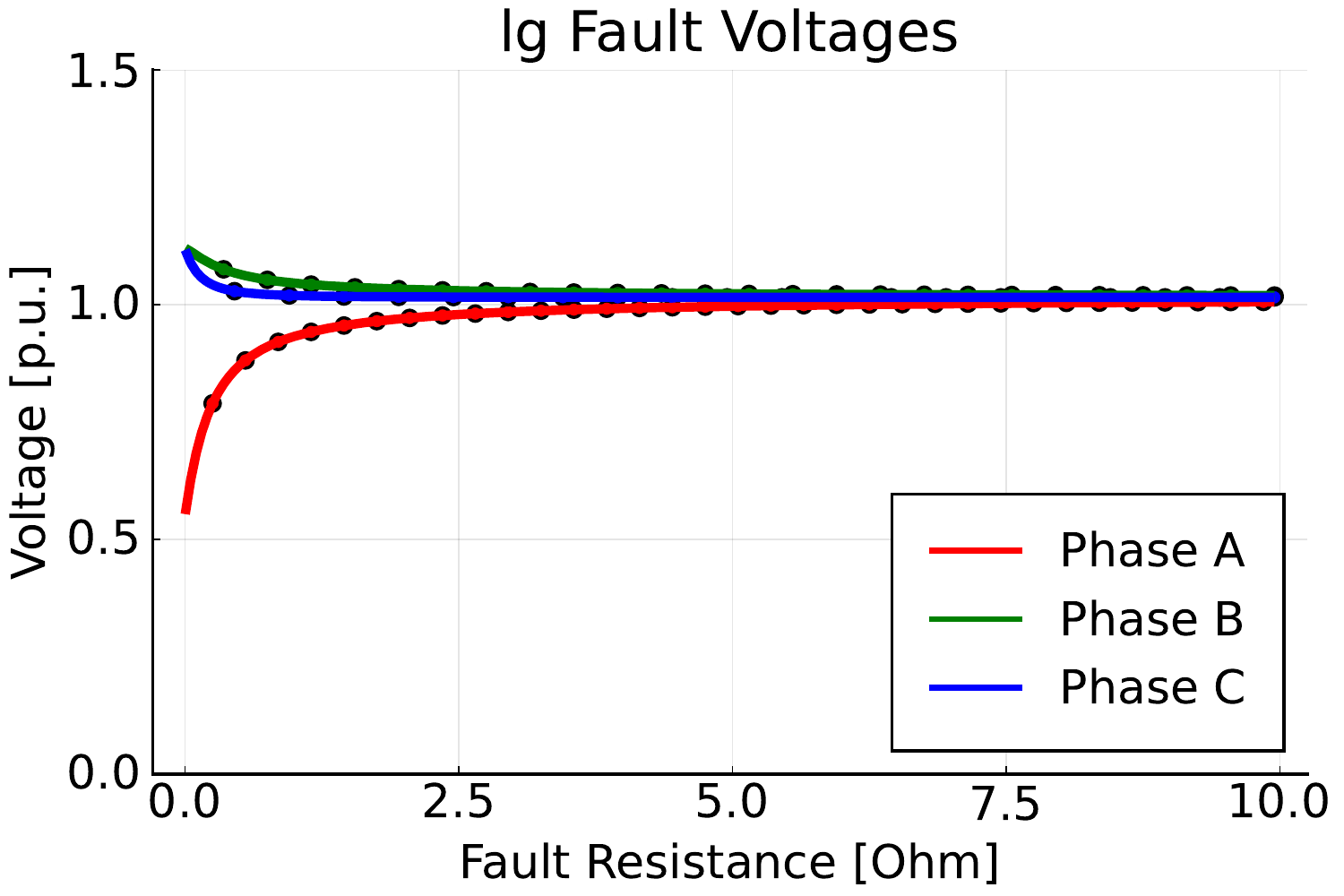}
    }
    \subfigure{
    \includegraphics[width=0.46\linewidth]{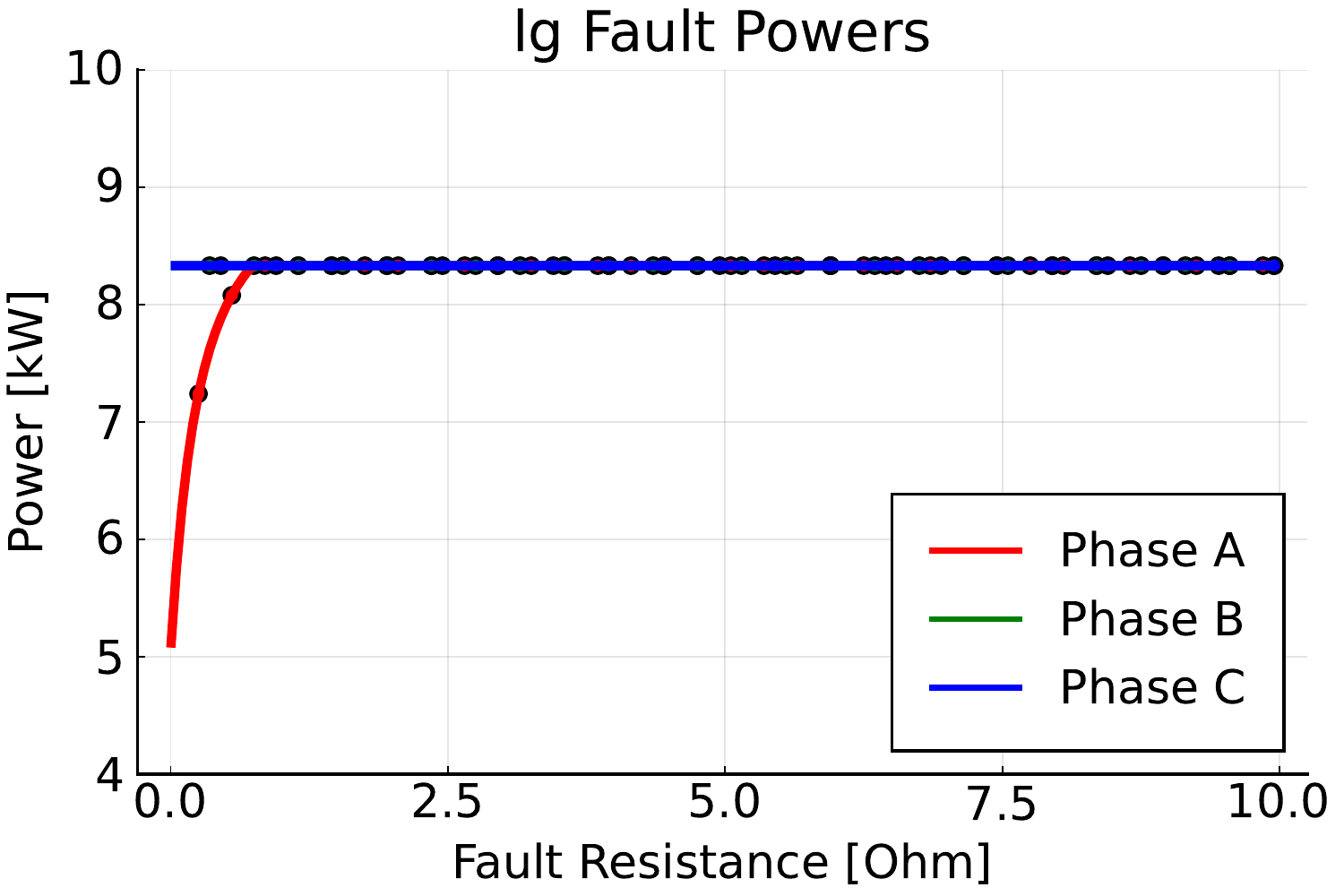}
    }
    \subfigure{
    \includegraphics[width=0.46\linewidth]{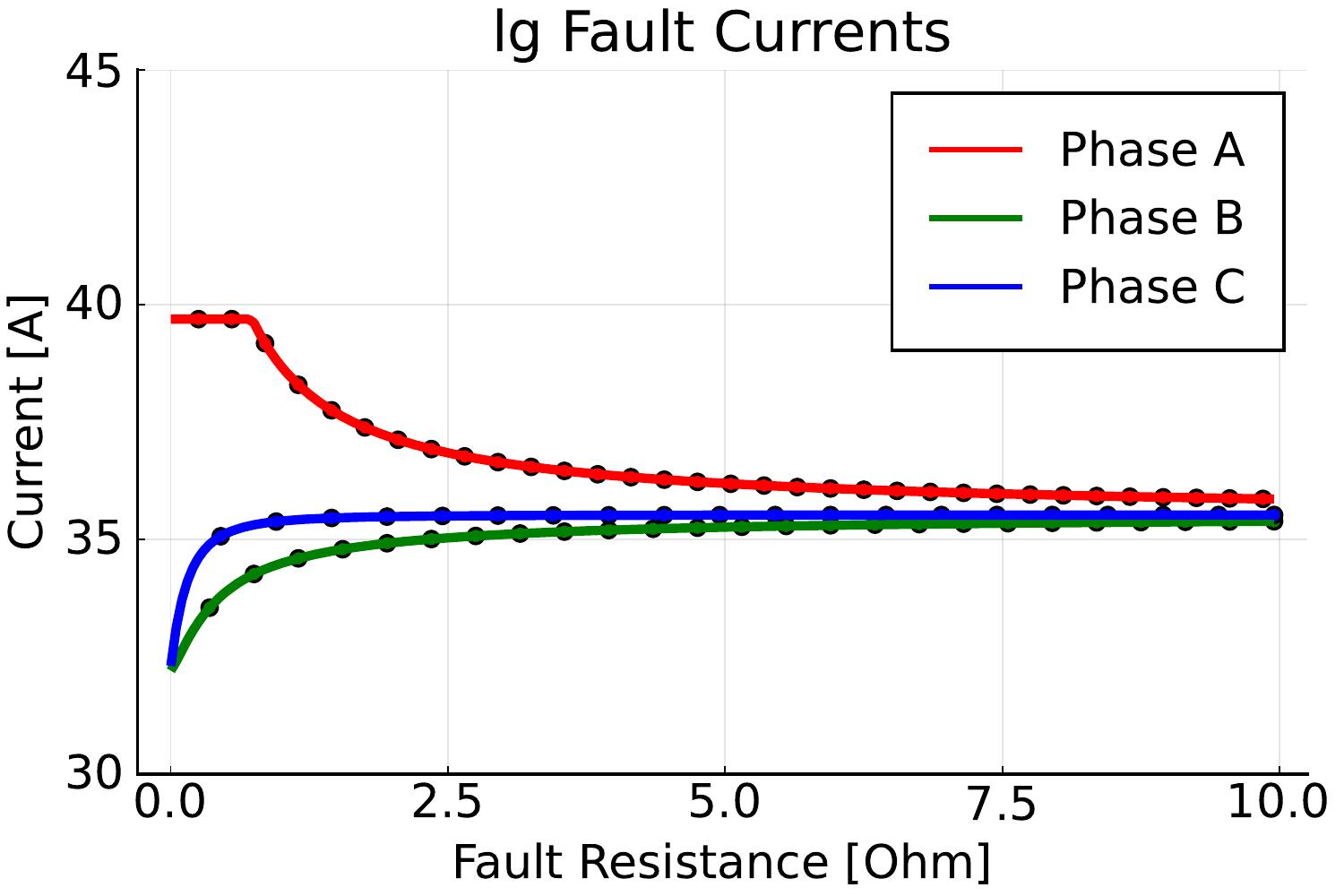}
    }
     \subfigure{
    \includegraphics[width=0.46\linewidth]{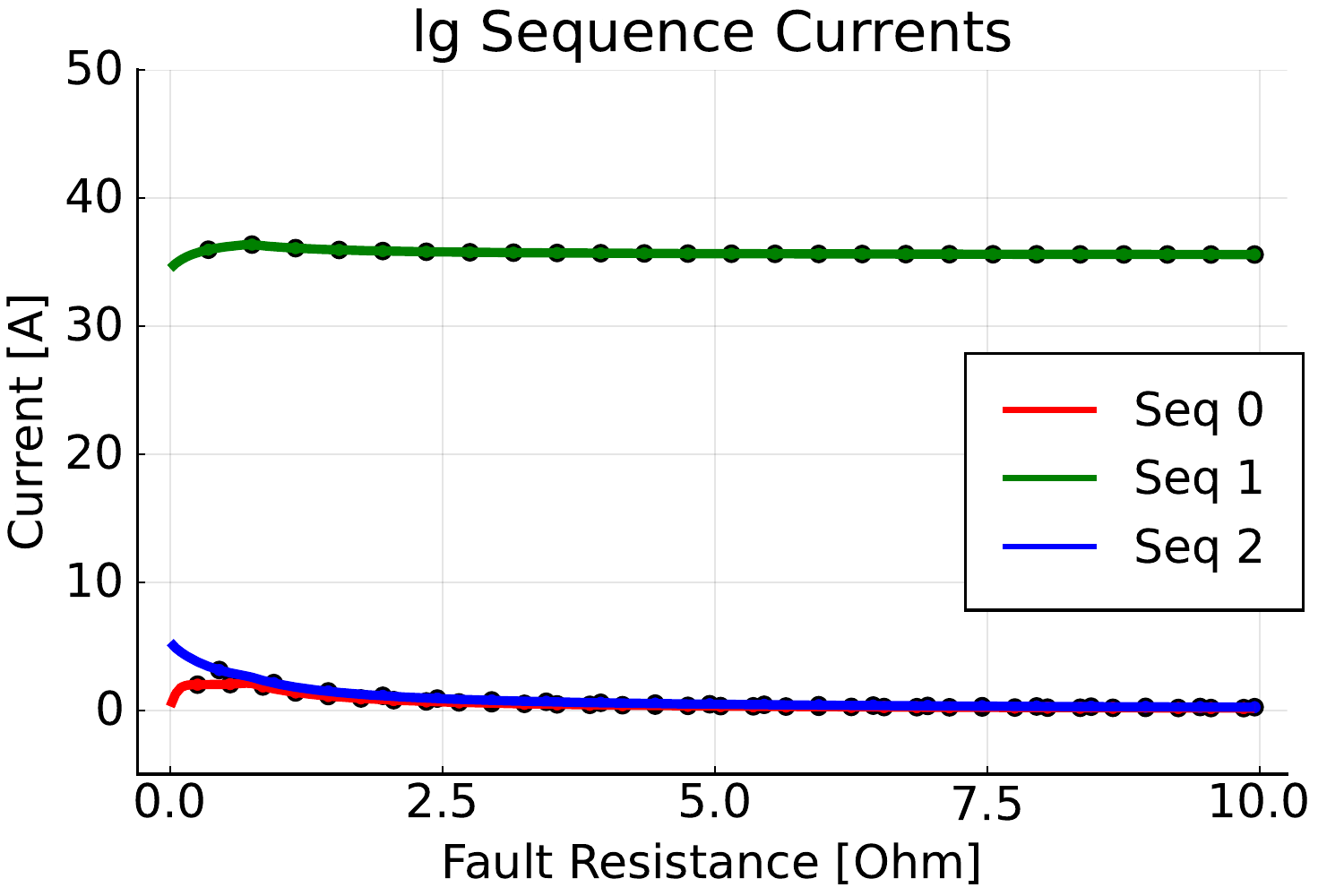}
    }
    \caption{Grid-Following Inverter -- Fault Currents [$A$] For Line-Ground Fault.}
    \label{fig:grid-follow-line-ground}
\end{figure}

The performance of the grid-following inverter model under line-to-line, single line-to-ground, and 3-phase faults was evaluated; the results are presented in Figs.~\ref{fig:grid-follow-line-line}-\ref{fig:grid-follow-line-ground}.

It can be observed that the inverter current saturates when the voltage collapses, resulting in a drop in the supplied active power from the inverter; the unfaulted phases, however, maintain the active power output.
The experimental results presented in \cite{8673877} showed that most of the sequence currents injected by grid-following inverters were positive sequence currents, which is in agreement with the sequence currents supplied by the developed grid-following inverter model. 

\begin{figure}[!htbp]
    \subfigure{
    \includegraphics[width=0.46\linewidth]{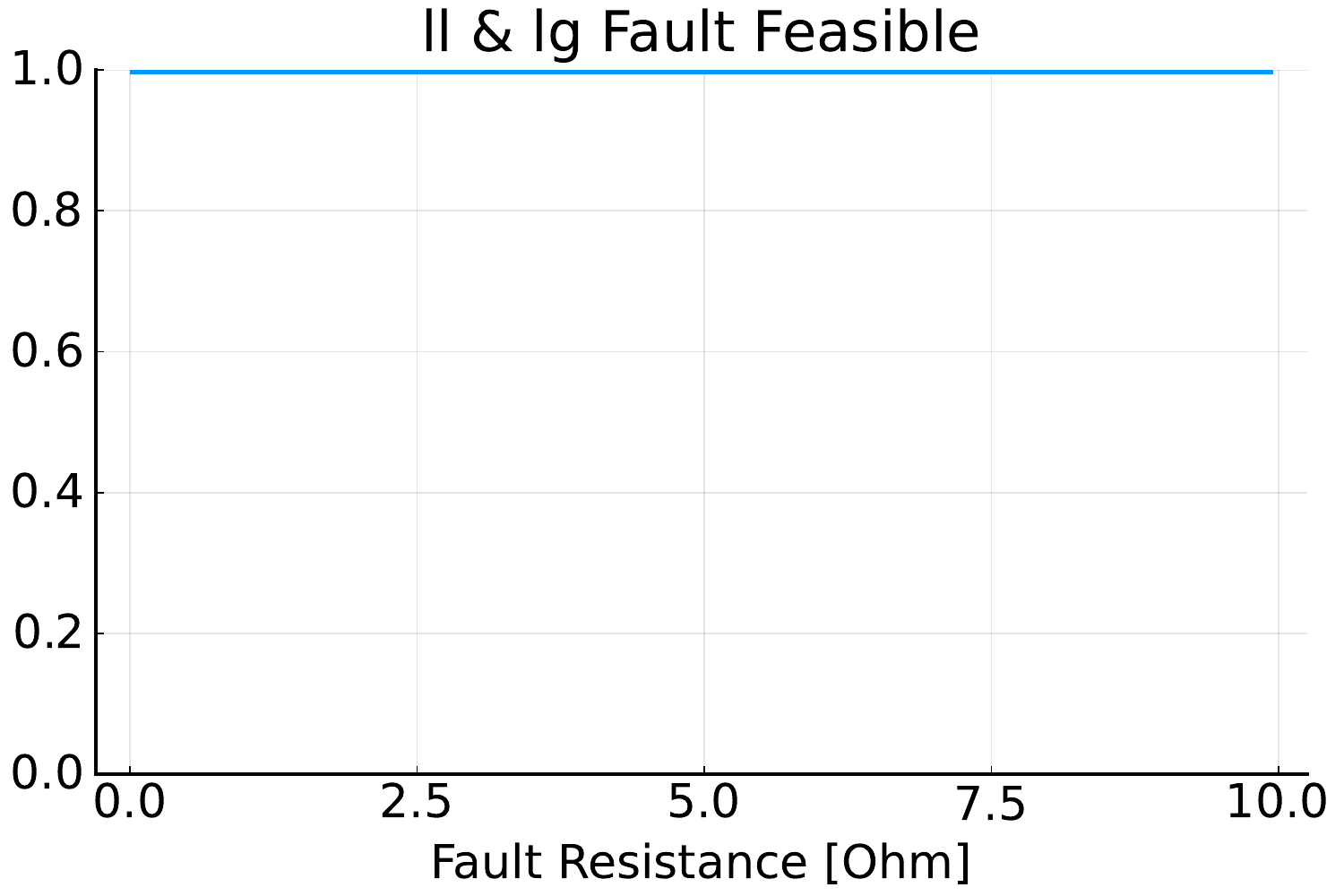}
    }
    \subfigure{
    \includegraphics[width=0.46\linewidth]{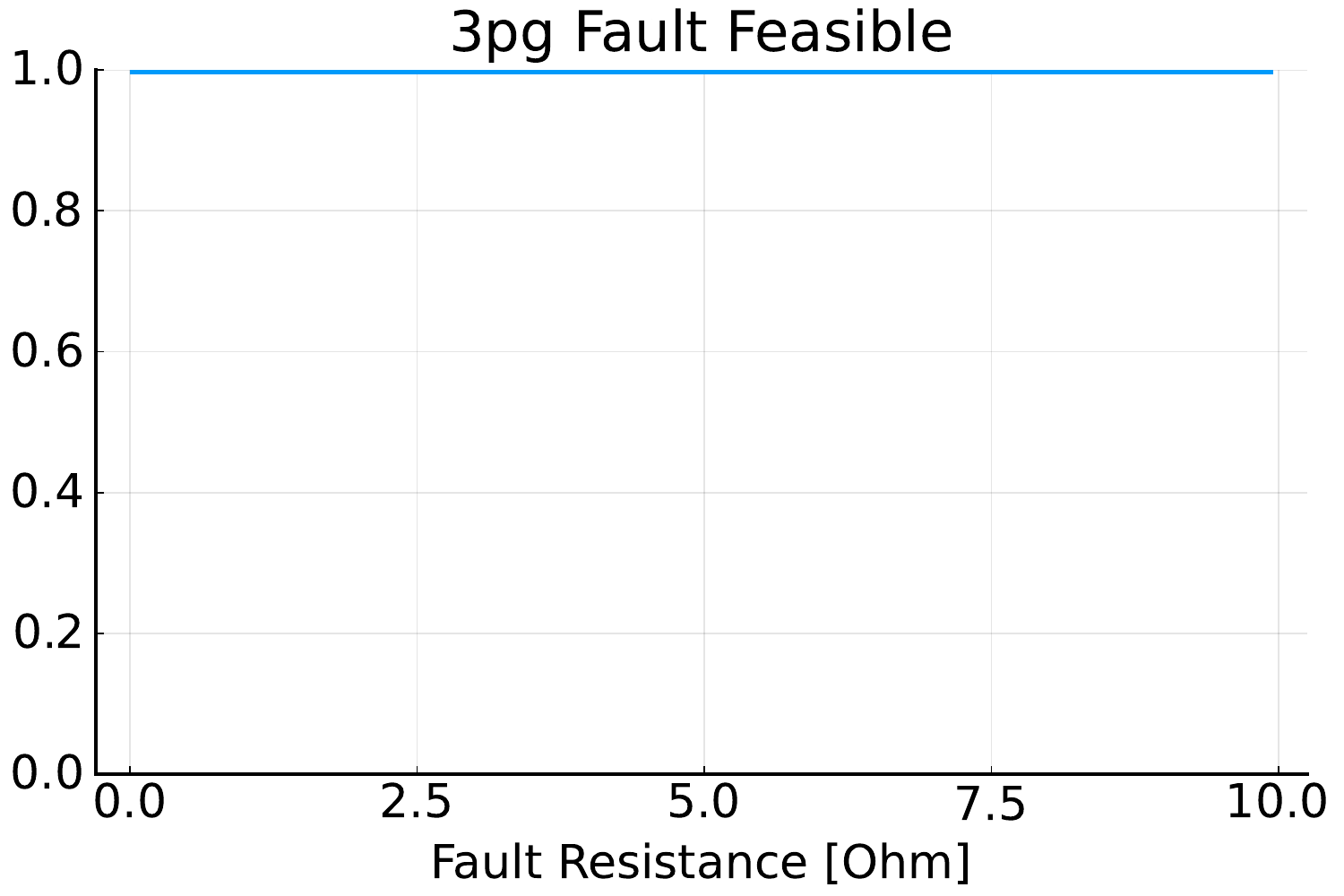}
    }
    \caption{Feasibility of the Grid-Following Inverter Under Faults.}
    \label{fig:grid-follow-feas}
\end{figure}

Fig.~\ref{fig:grid-follow-feas} demonstrates that the solver returned a feasible solution for each inverter over the faulted impedance sweep.

\subsection{Grid-Forming Inverter Model}
\normalsize \indent

\begin{figure}[!htbp]
    \subfigure{
    \includegraphics[width=0.46\linewidth]{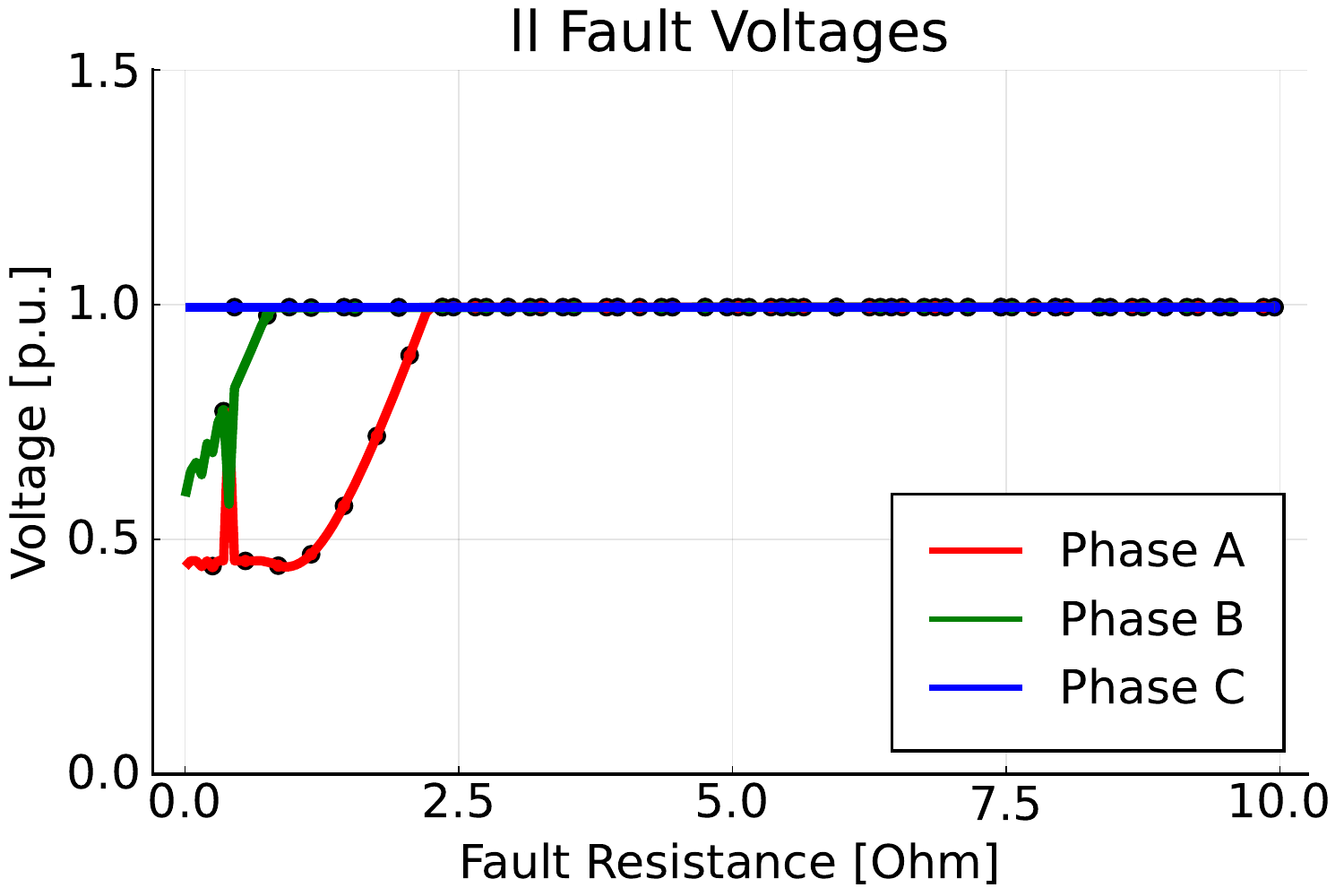}
    }
    \subfigure{
    \includegraphics[width=0.46\linewidth]{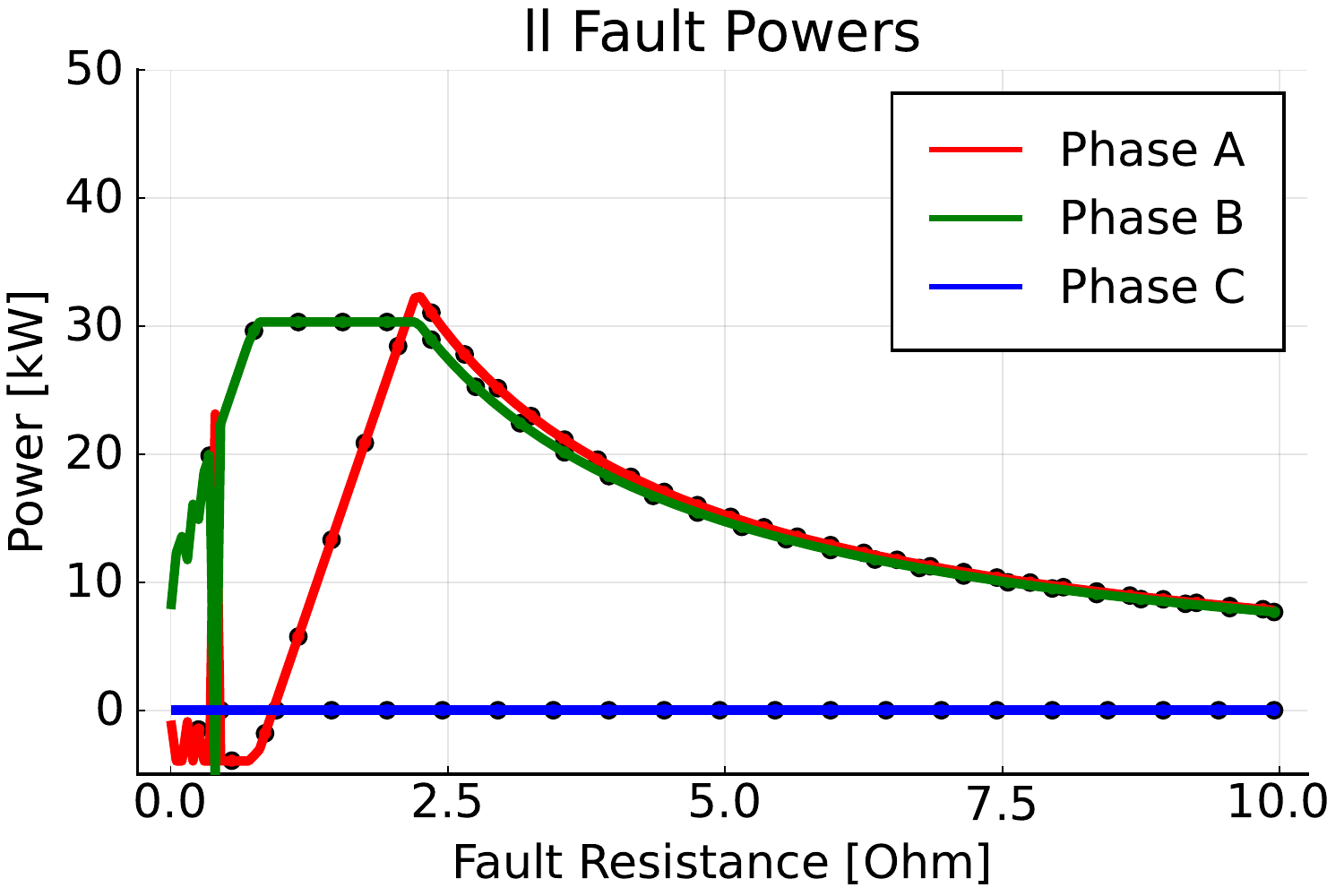}
    }
    \subfigure{
    \includegraphics[width=0.46\linewidth]{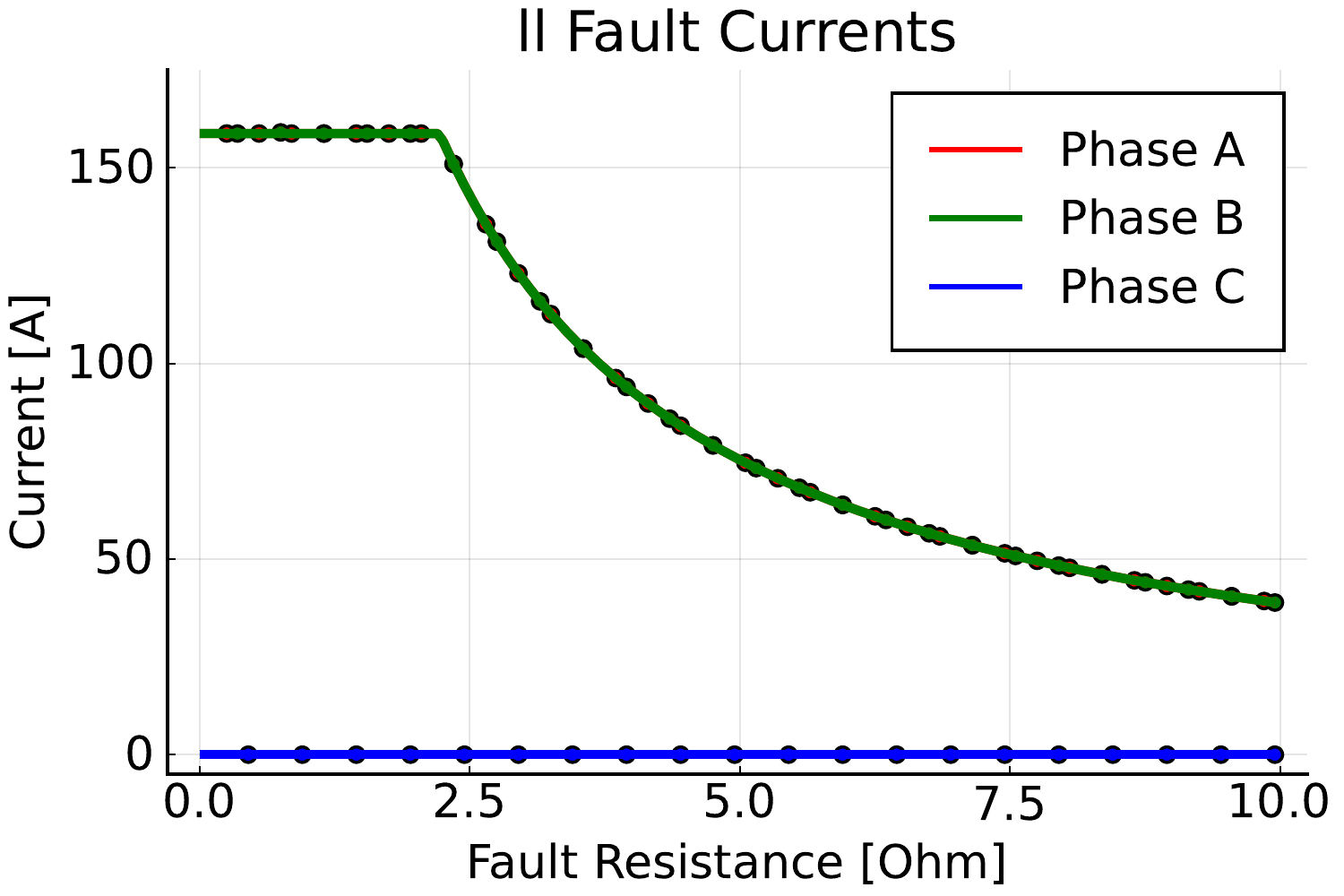}
    }
     \subfigure{
    \includegraphics[width=0.46\linewidth]{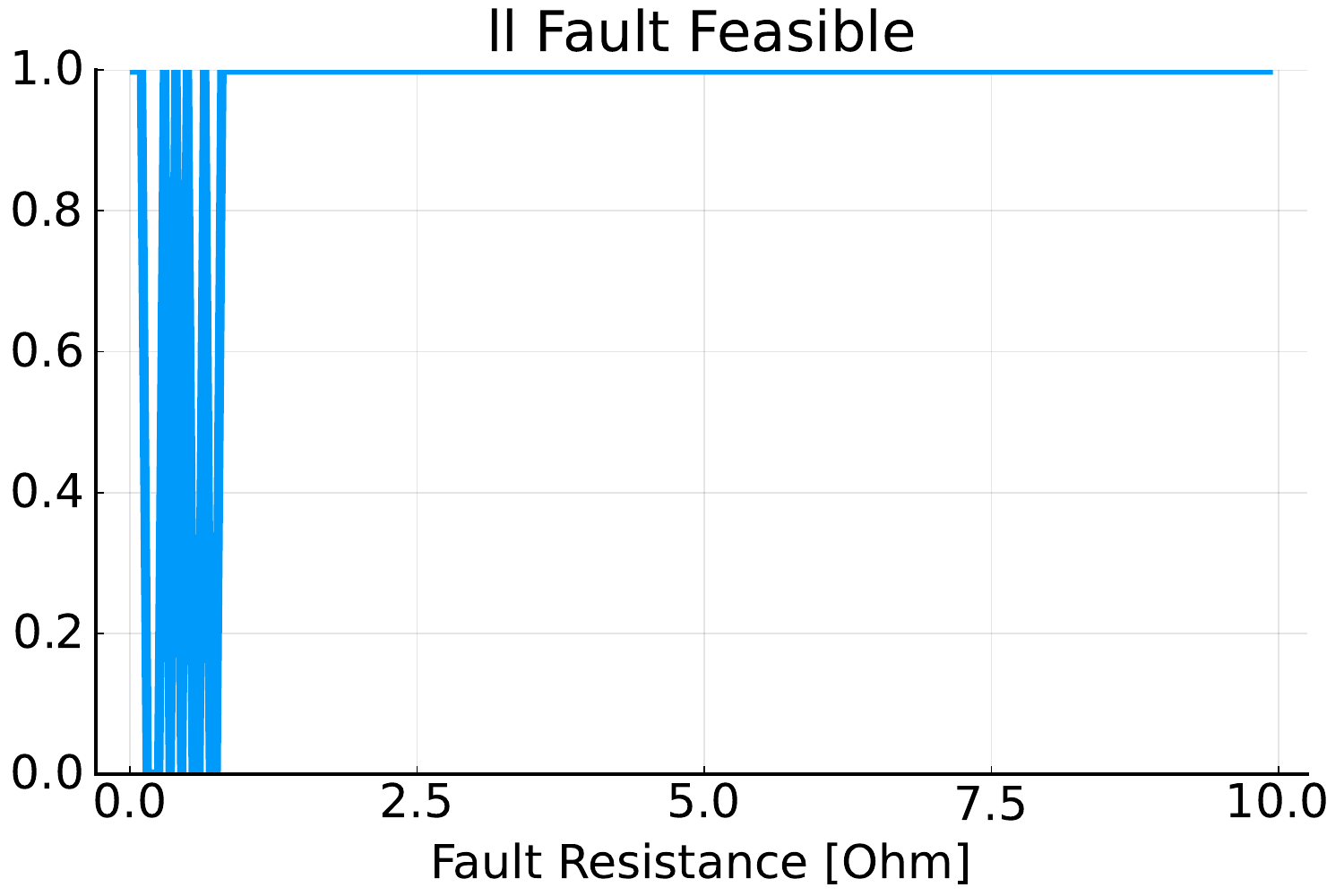}
    }
    \caption{Simple Grid-Forming -- Fault Currents [$A$] For Line-Line Fault.}
    \label{fig:grid-forming-line-line}
\end{figure}

\begin{figure}[!htbp]
    \subfigure{
    \includegraphics[width=0.46\linewidth]{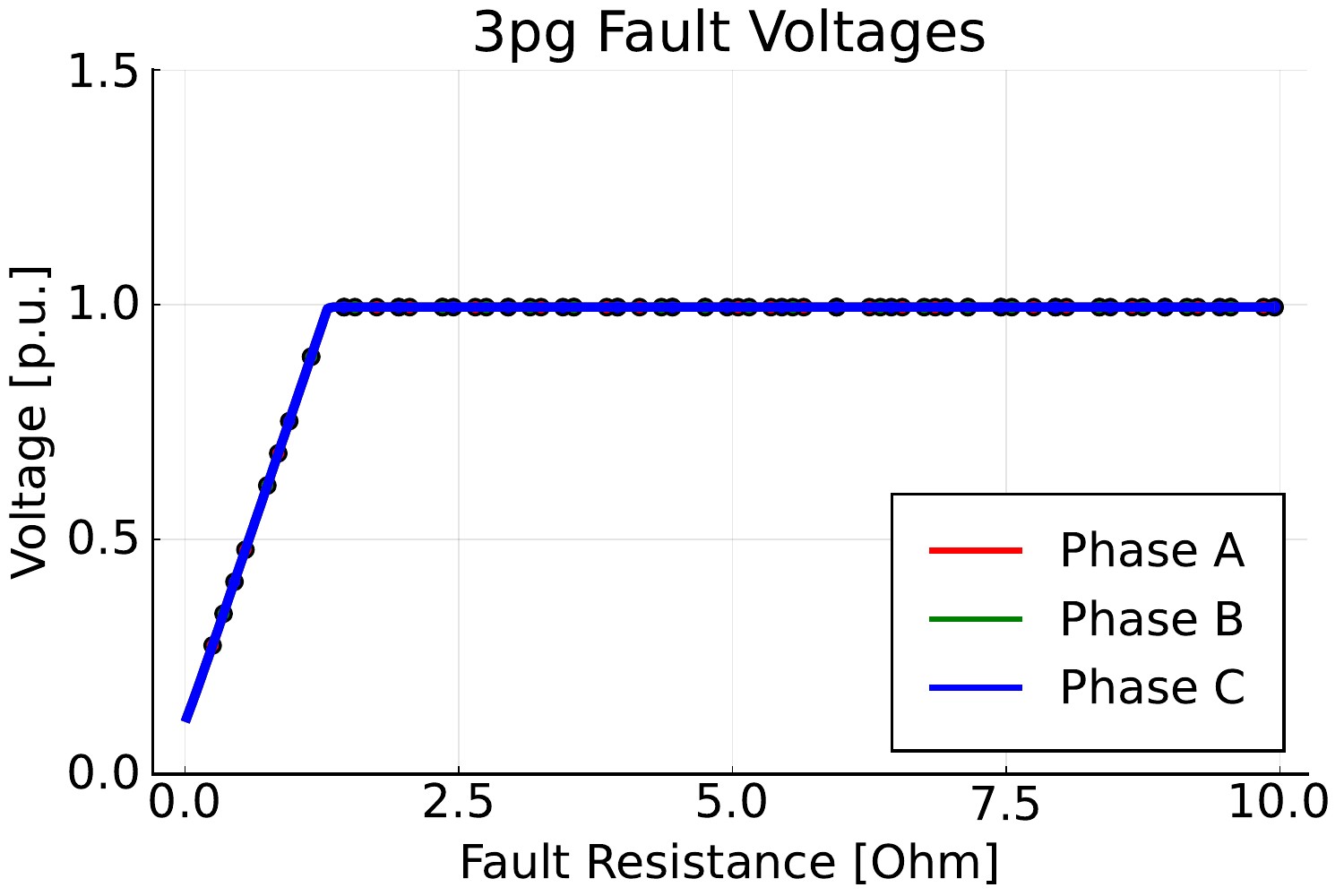}
    }
    \subfigure{
    \includegraphics[width=0.46\linewidth]{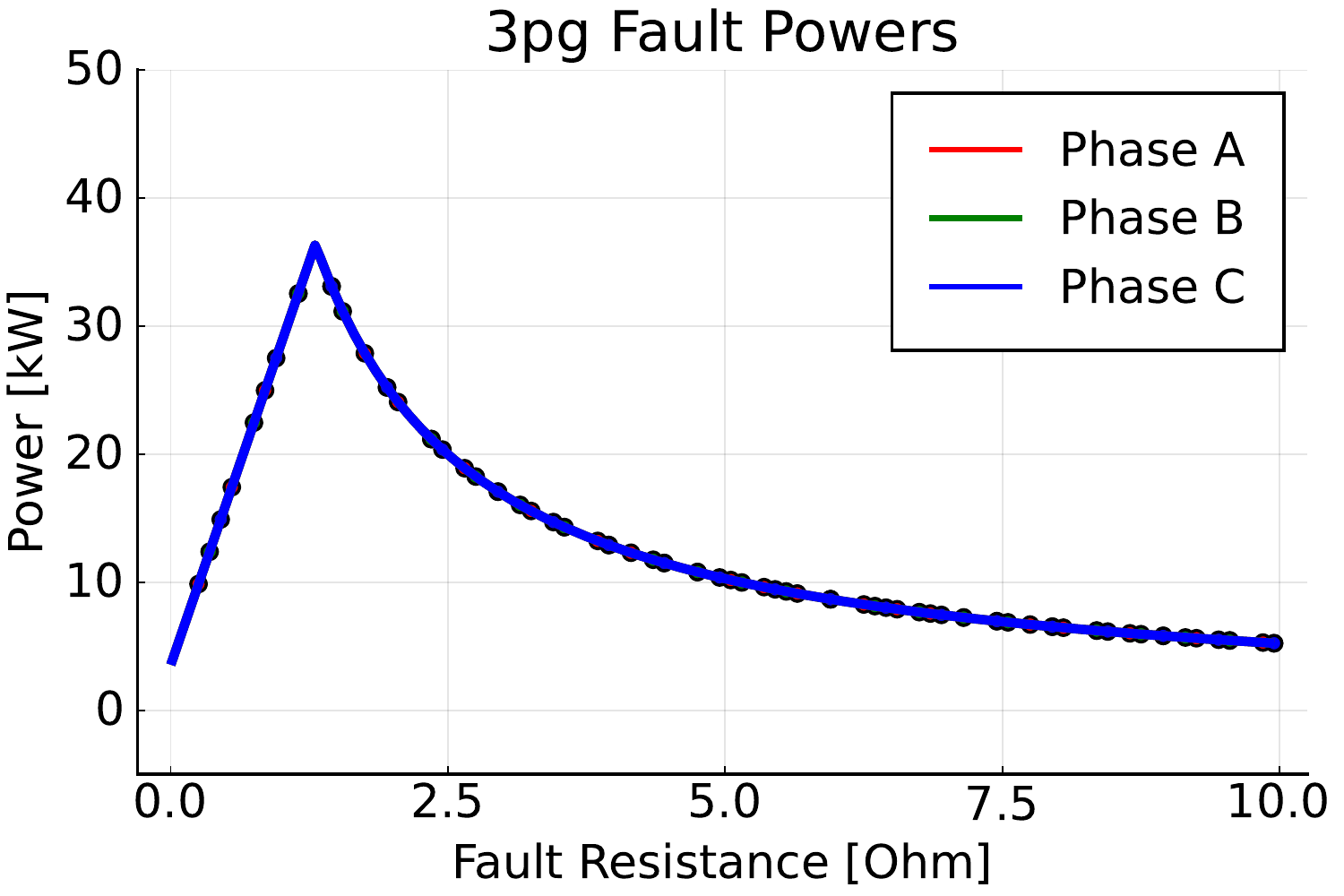}
    }
    \subfigure{
    \includegraphics[width=0.46\linewidth]{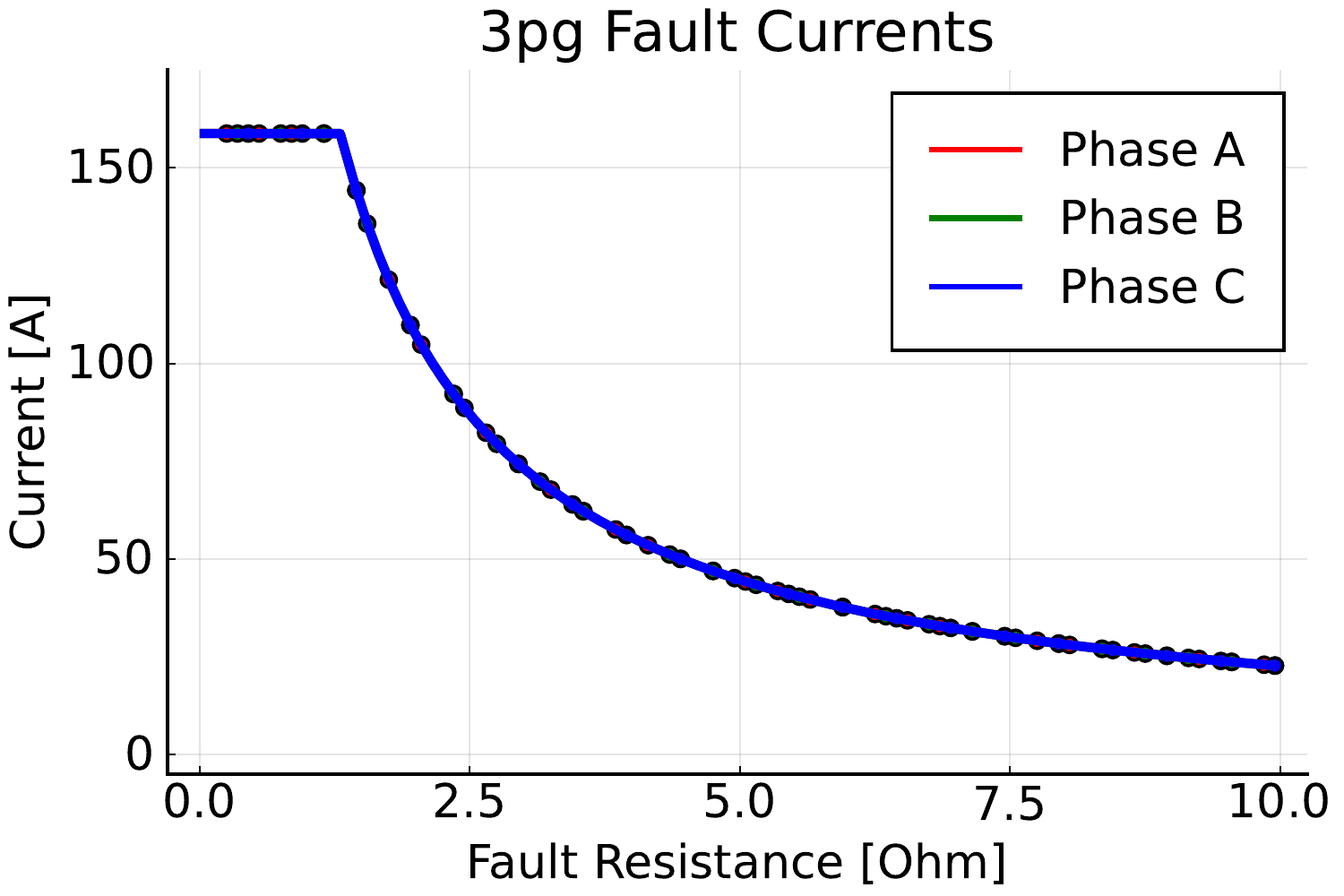}
    }
     \subfigure{
    \includegraphics[width=0.46\linewidth]{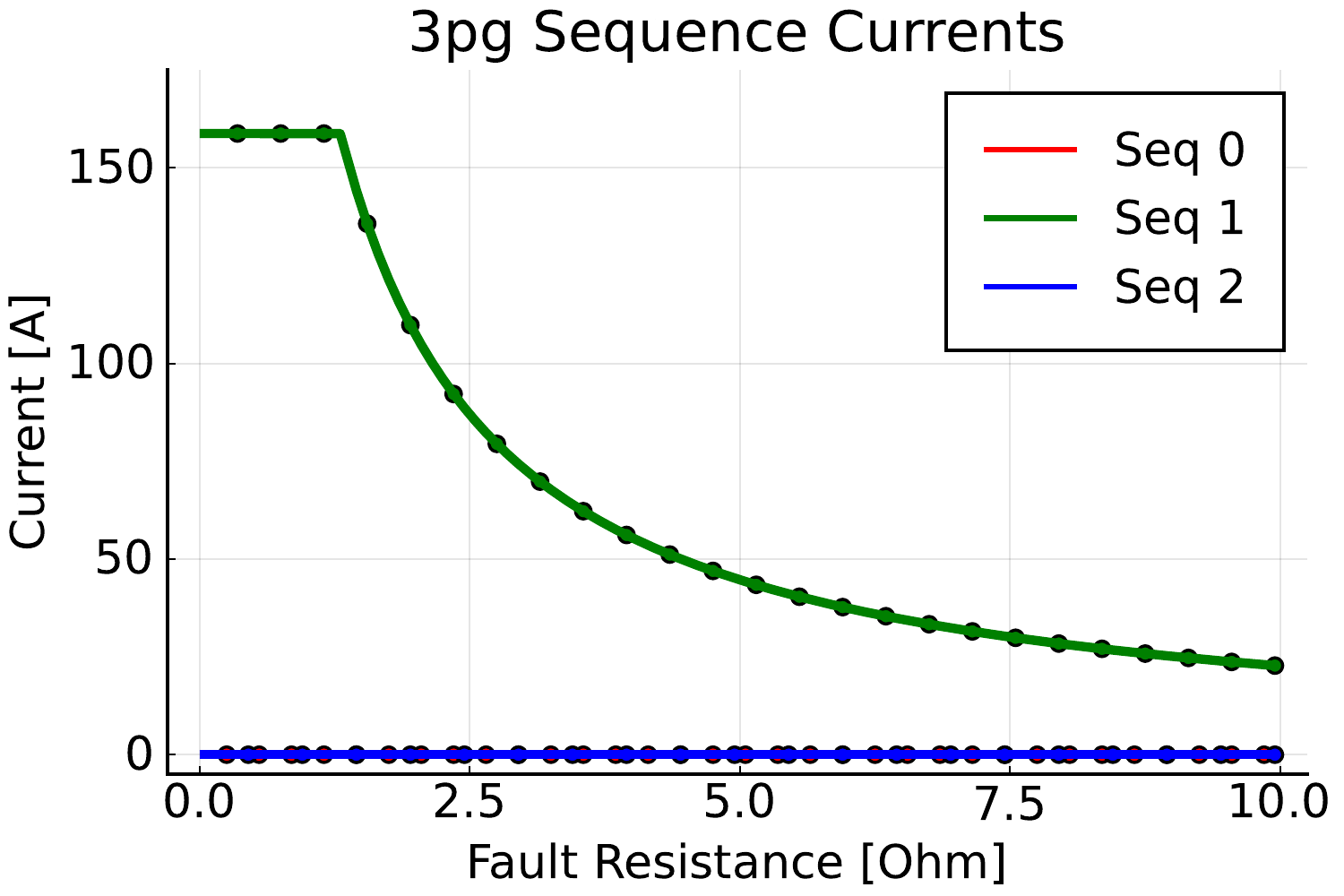}
    }
    \caption{Simple Grid-Forming -- Fault Currents [$A$] For 3-Phase Fault.}
    \label{fig:grid-forming-3ph}
\end{figure}

\begin{figure}[!htbp]
    \subfigure{
    \includegraphics[width=0.46\linewidth]{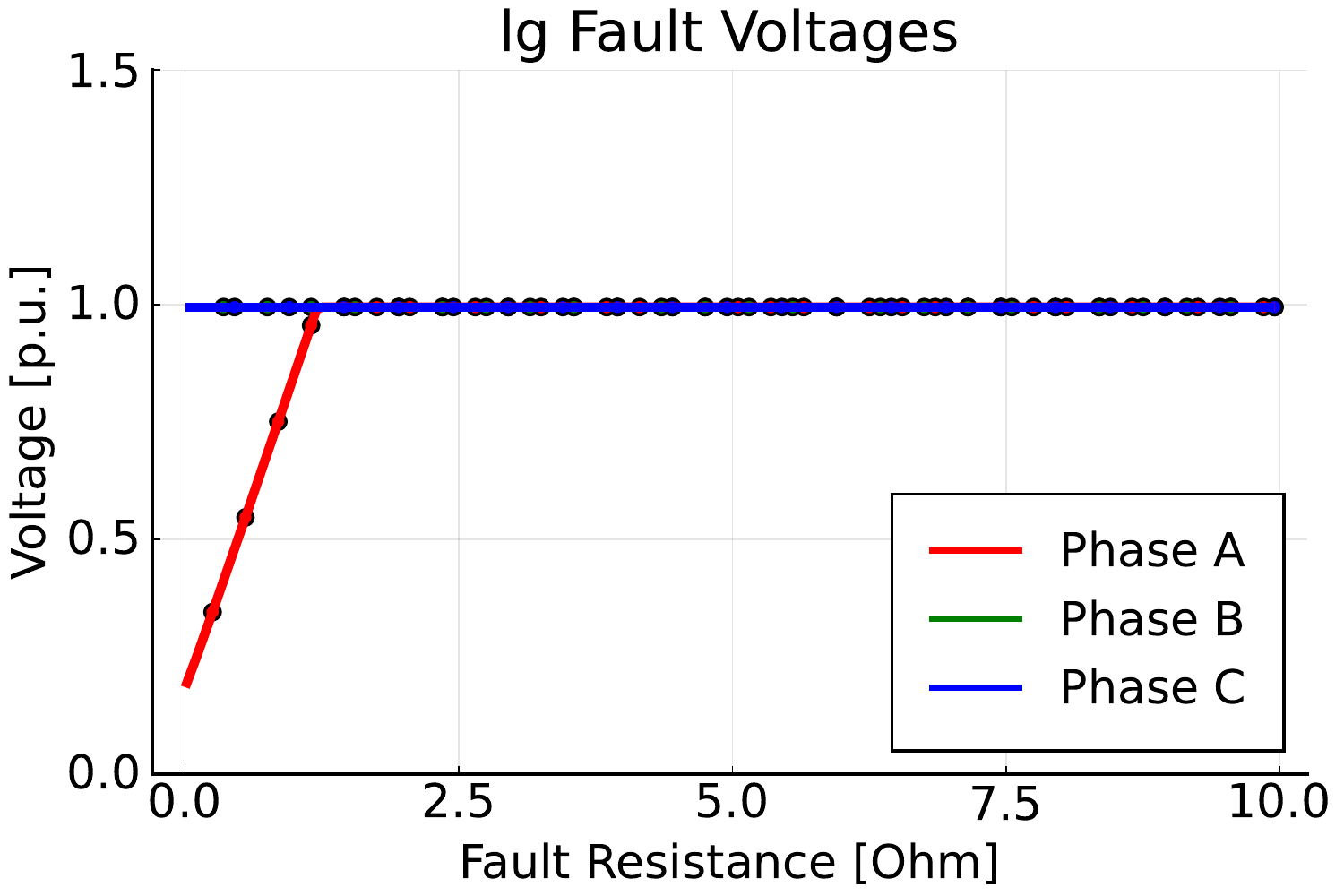}
    }
    \subfigure{
    \includegraphics[width=0.46\linewidth]{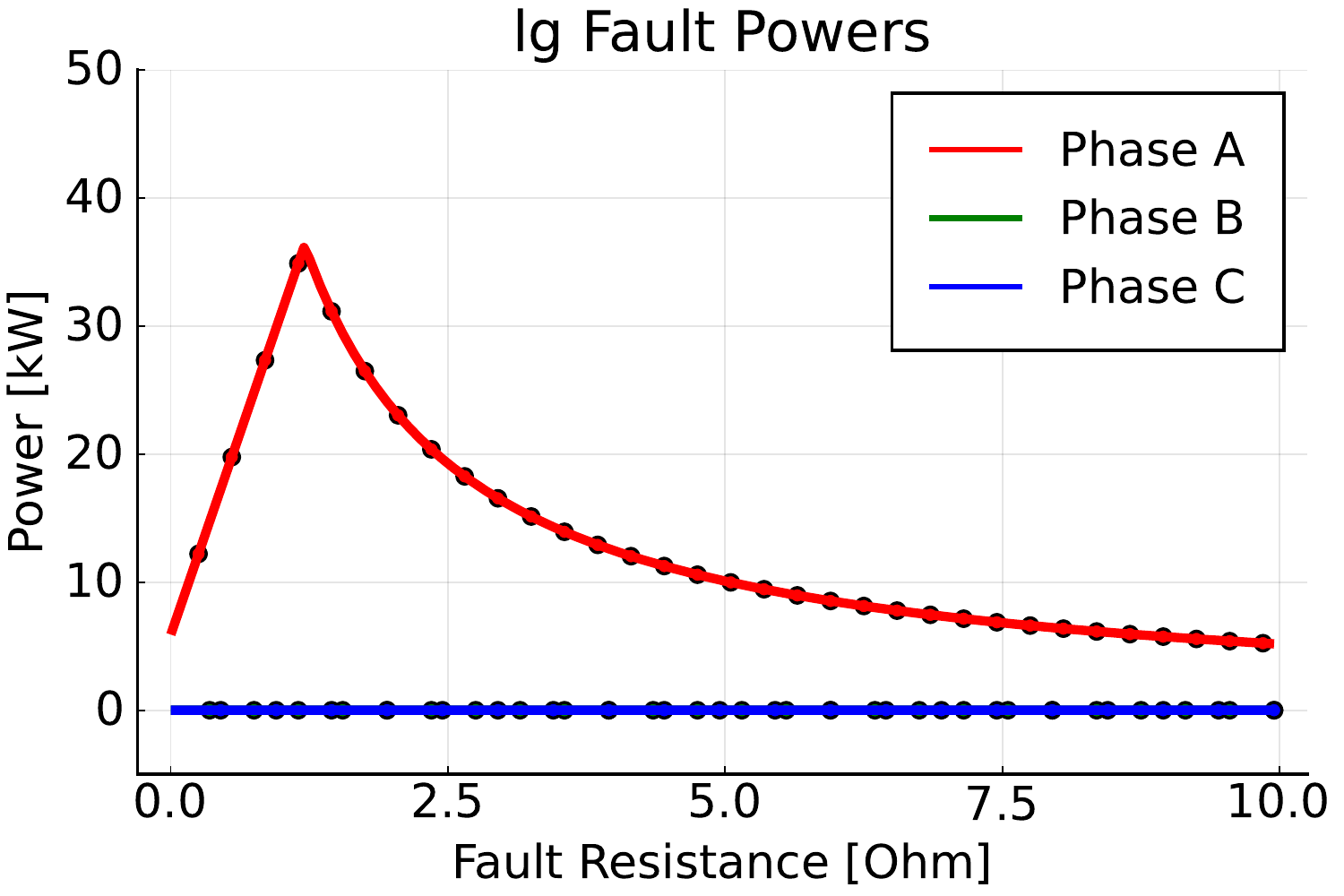}
    }
    \subfigure{
    \includegraphics[width=0.46\linewidth]{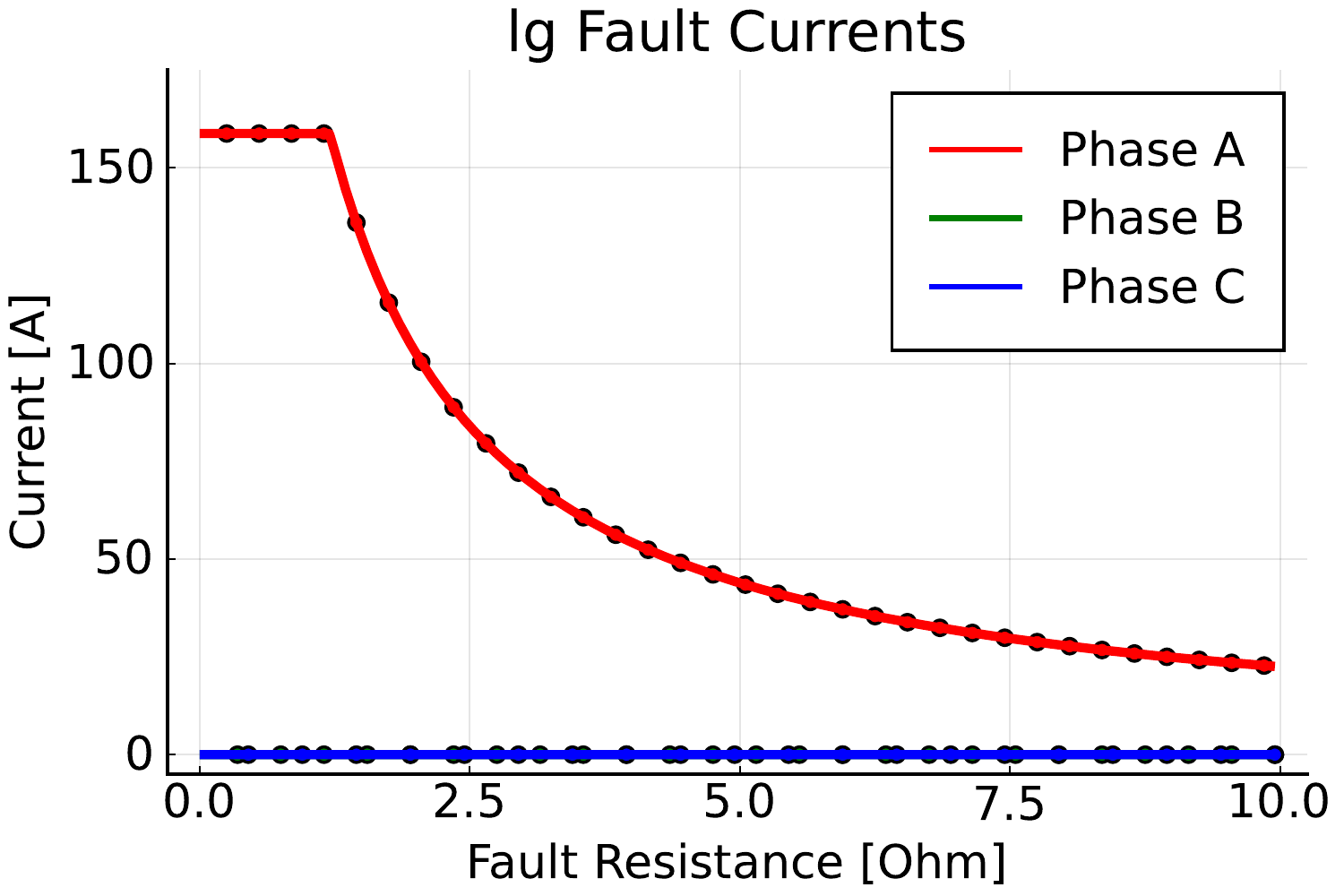}
    }
     \subfigure{
    \includegraphics[width=0.46\linewidth]{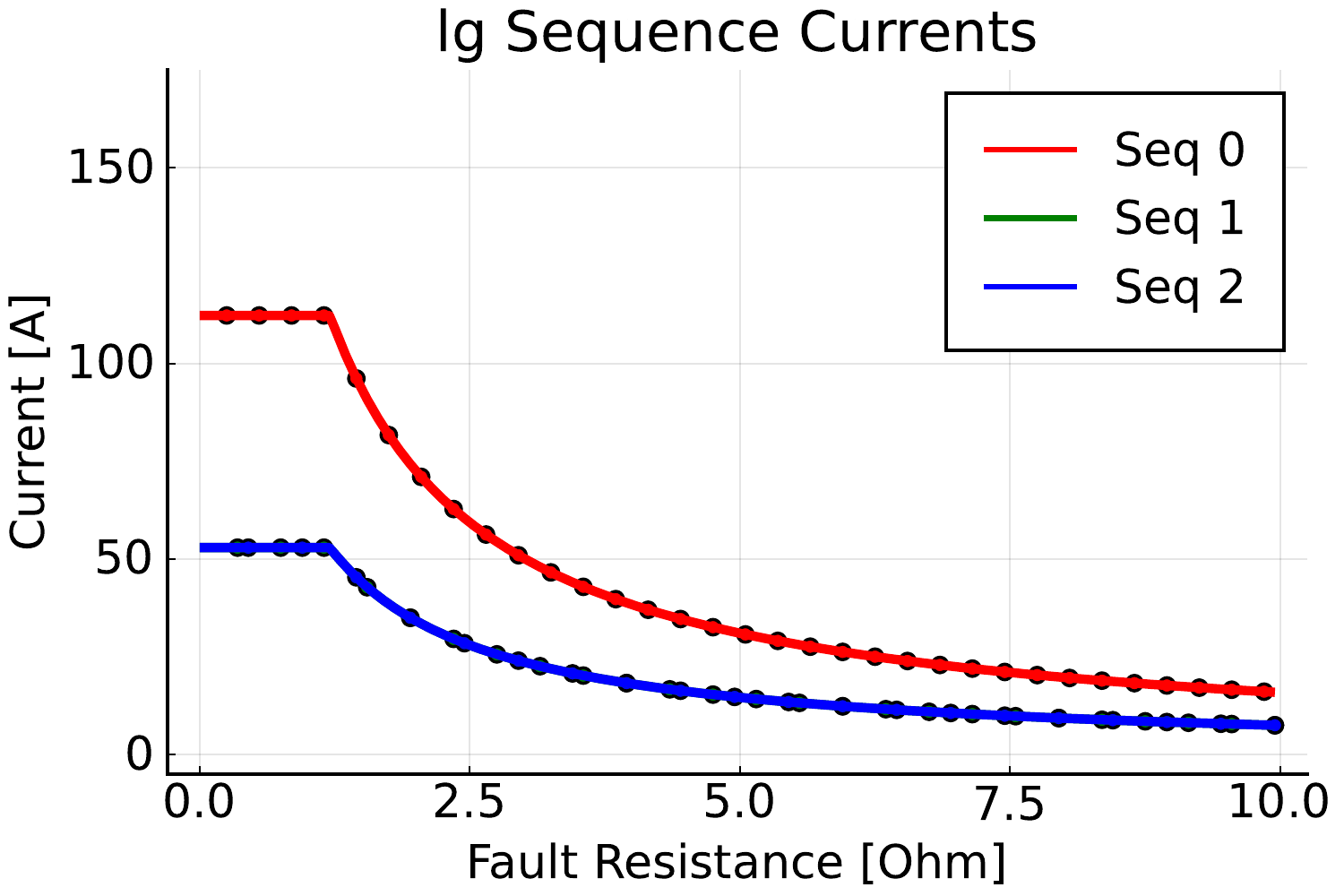}
    }
    \caption{Simple Grid-Forming -- Fault Currents [$A$] For Line-Ground Fault.}
    \label{fig:grid-forming-line-ground}
\end{figure}

\begin{figure}[!htbp]
    \subfigure{
    \includegraphics[width=0.46\linewidth]{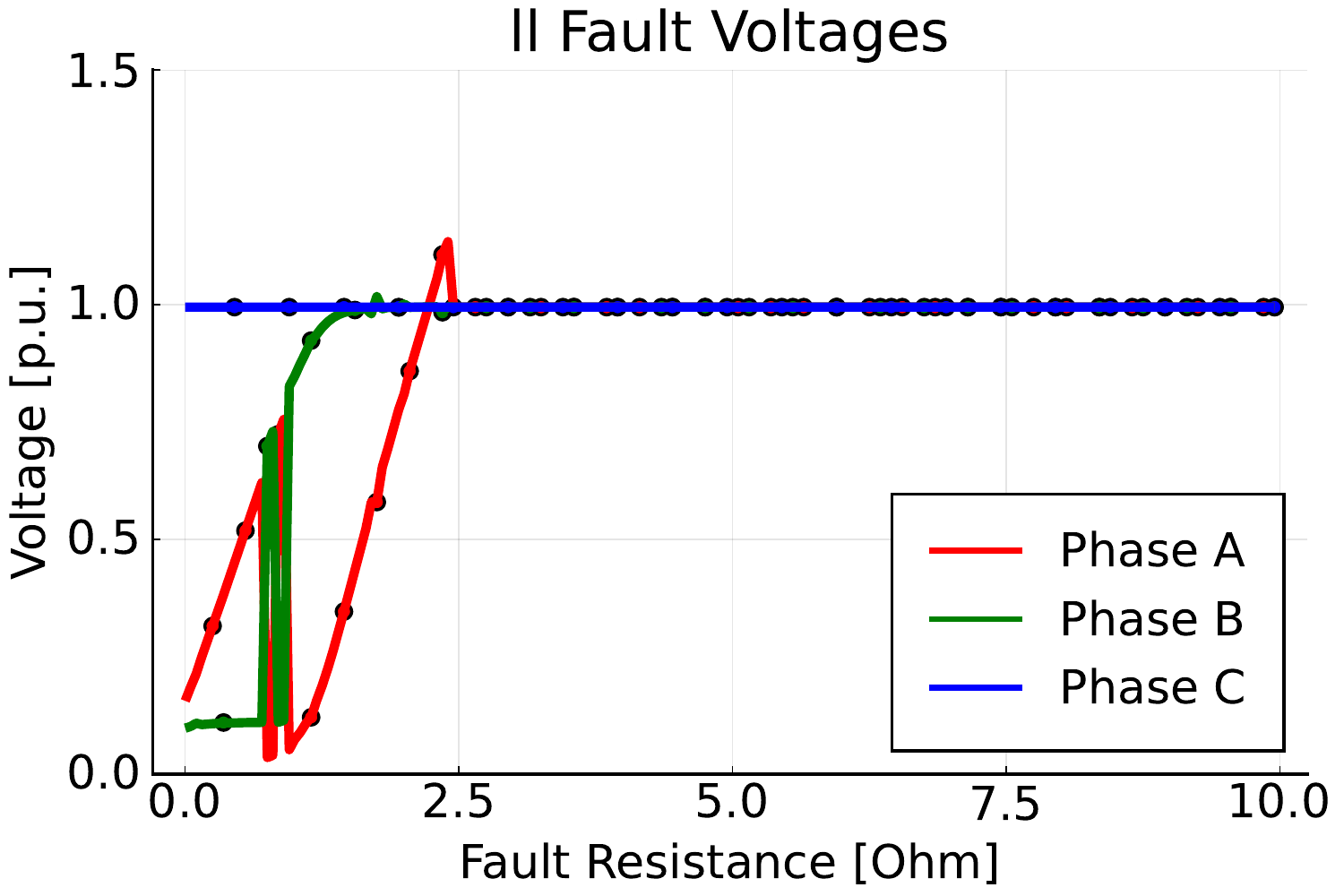}
    }
    \subfigure{
    \includegraphics[width=0.46\linewidth]{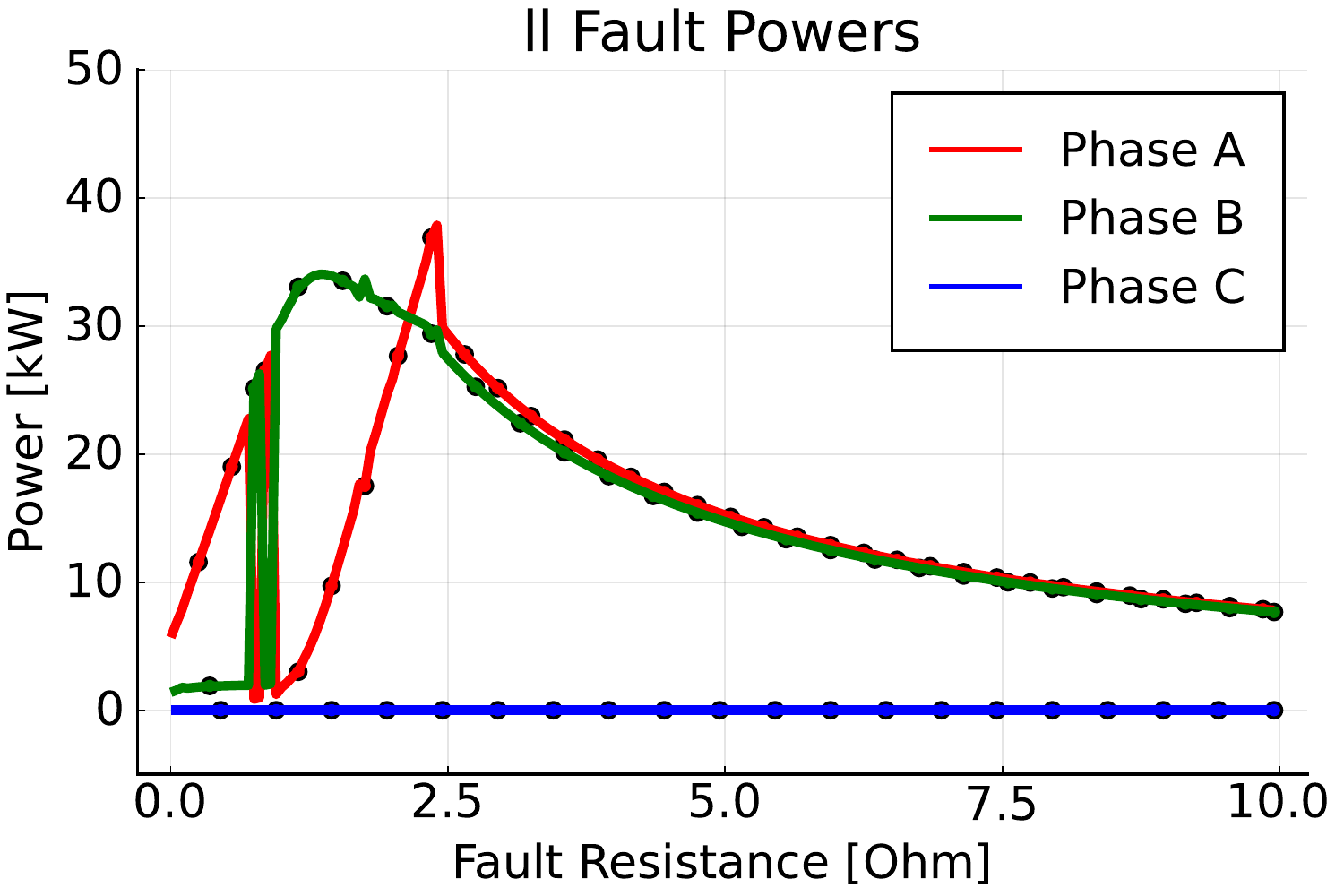}
    }
    \subfigure{
    \includegraphics[width=0.46\linewidth]{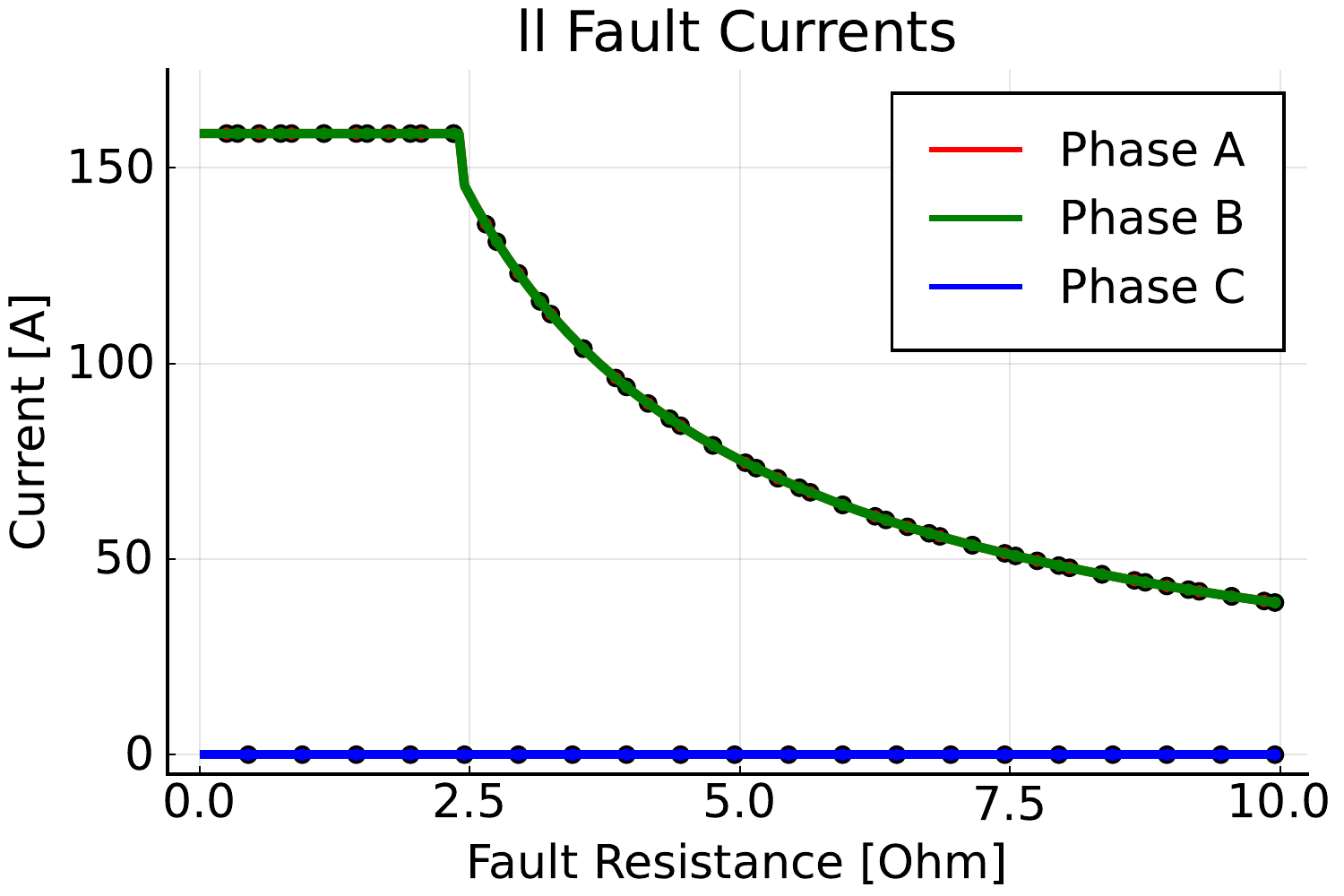}
    }
     \subfigure{
    \includegraphics[width=0.46\linewidth]{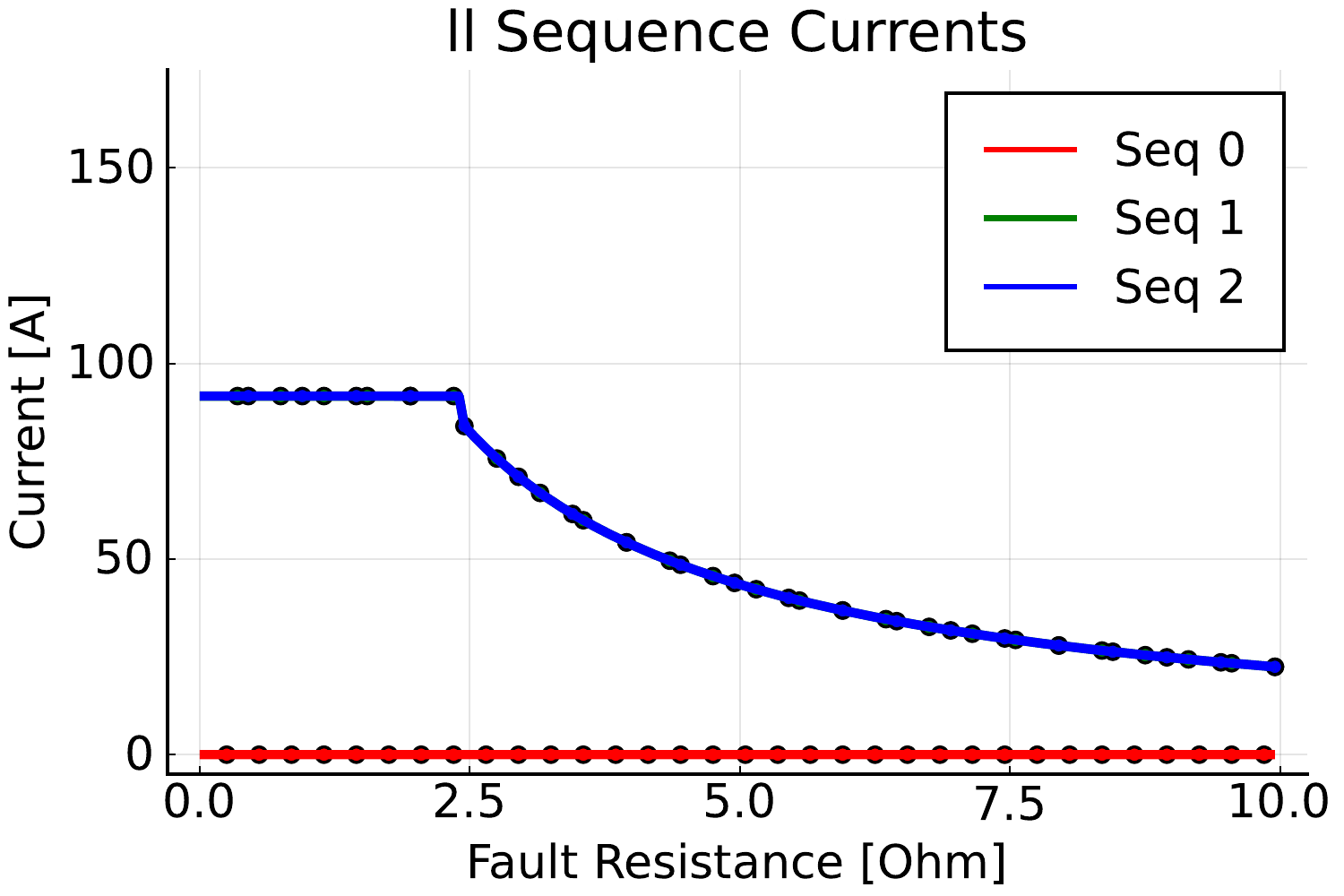}
    }
    \caption{Complex Grid-Forming Fault Currents [$A$] For Line-Line Fault.}
    \label{fig:grid-forming-line-line_2}
\end{figure}

\begin{figure}[!htbp]
    \subfigure{
    \includegraphics[width=0.46\linewidth]{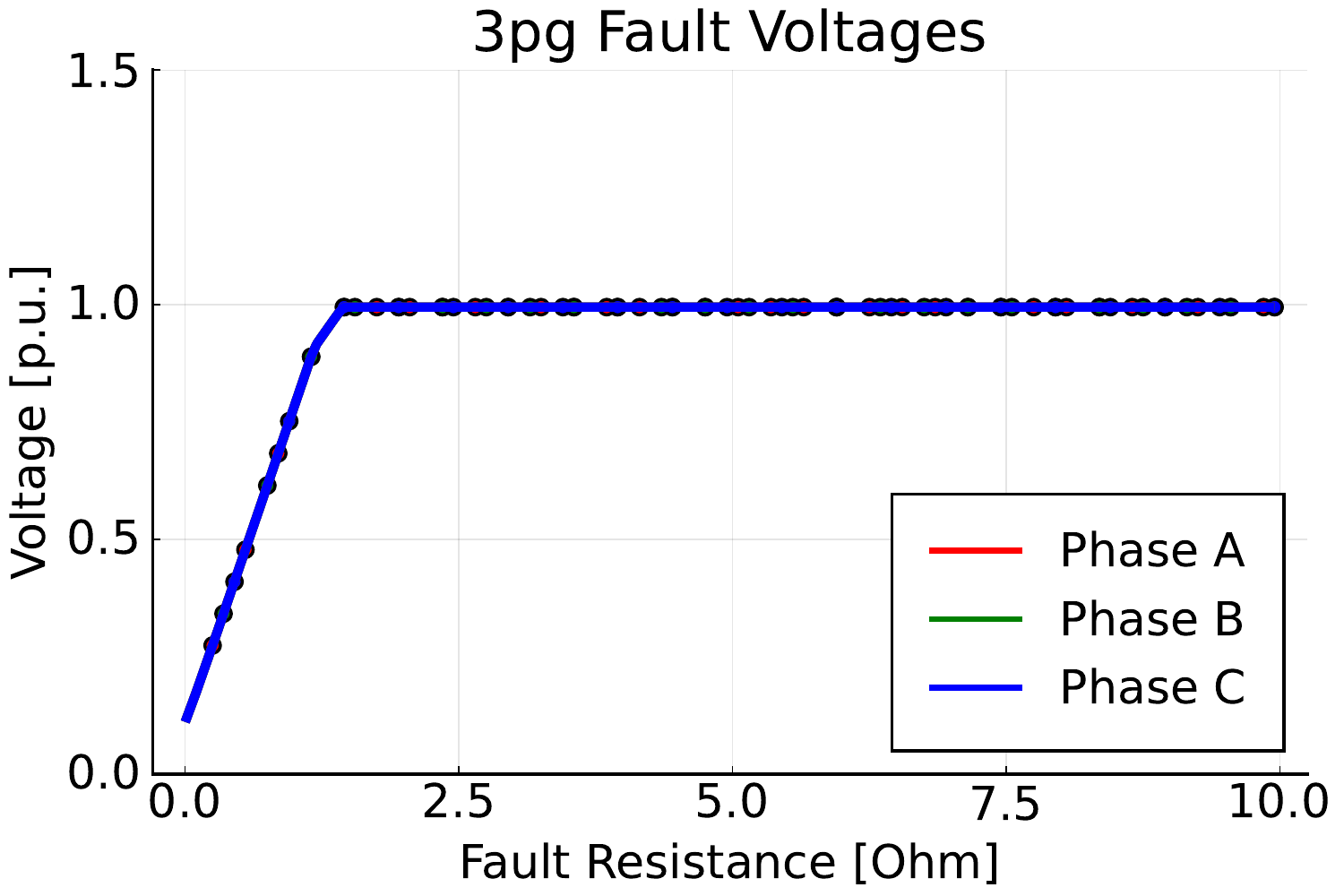}
    }
    \subfigure{
    \includegraphics[width=0.46\linewidth]{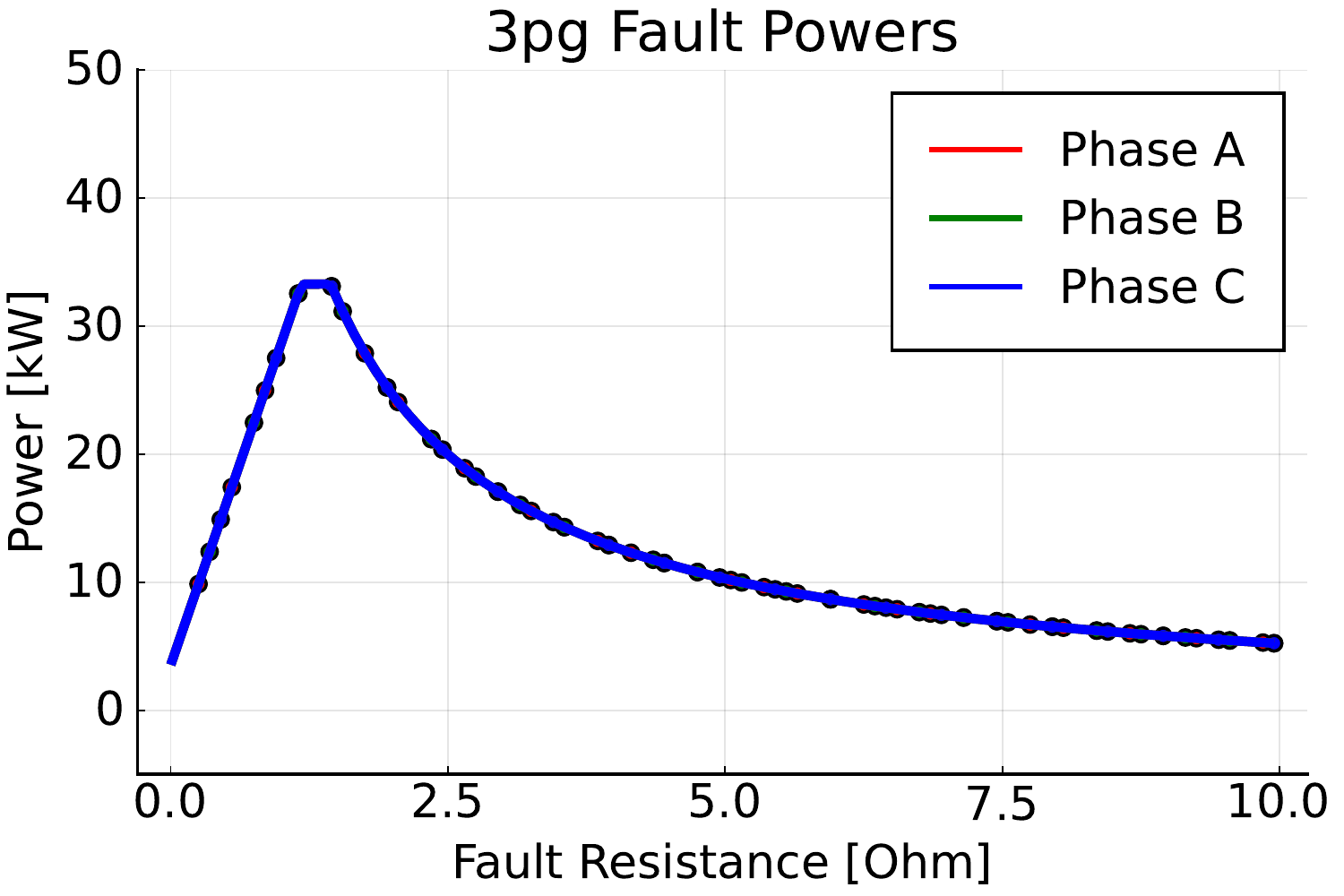}
    }
    \subfigure{
    \includegraphics[width=0.46\linewidth]{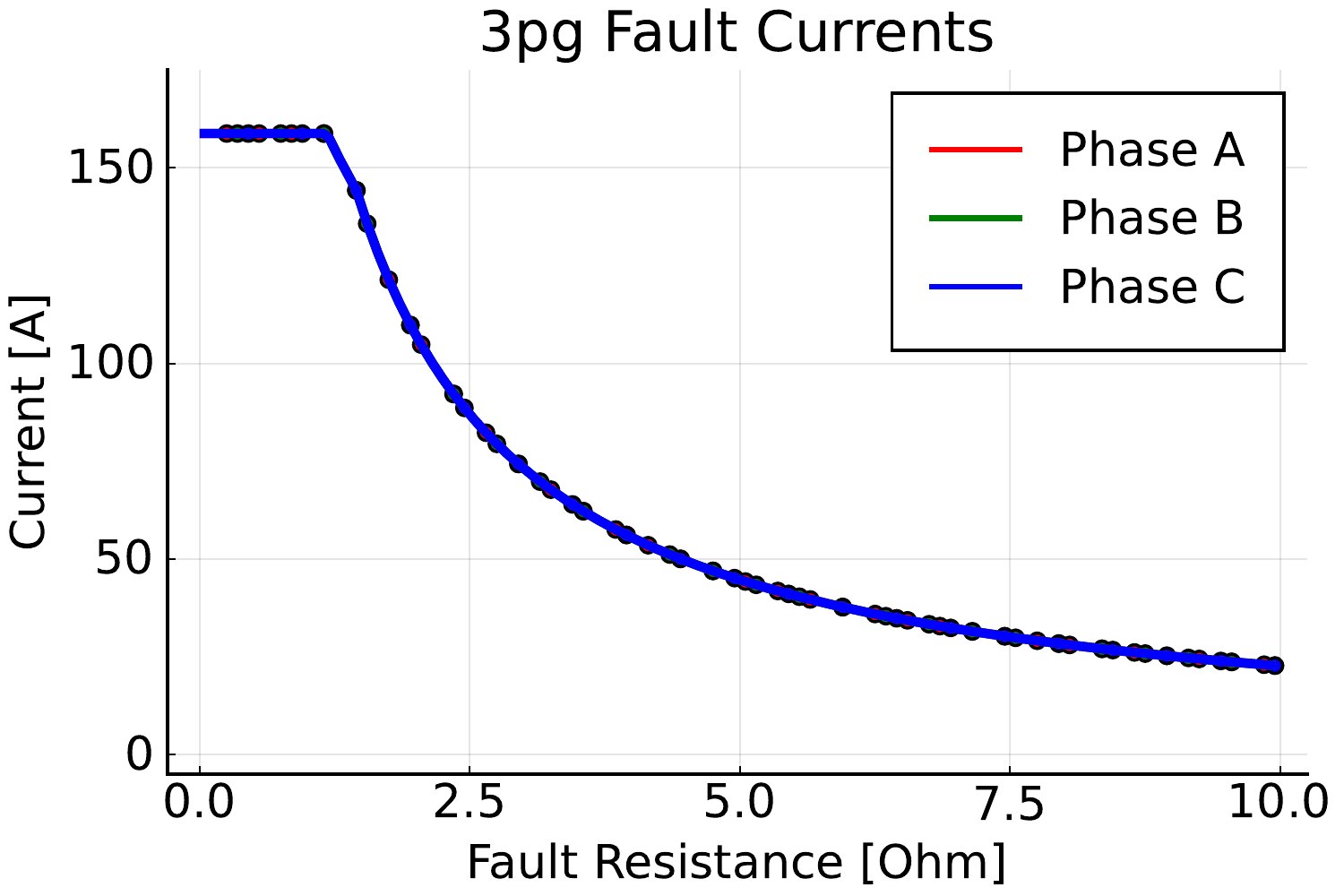}
    }
     \subfigure{
    \includegraphics[width=0.46\linewidth]{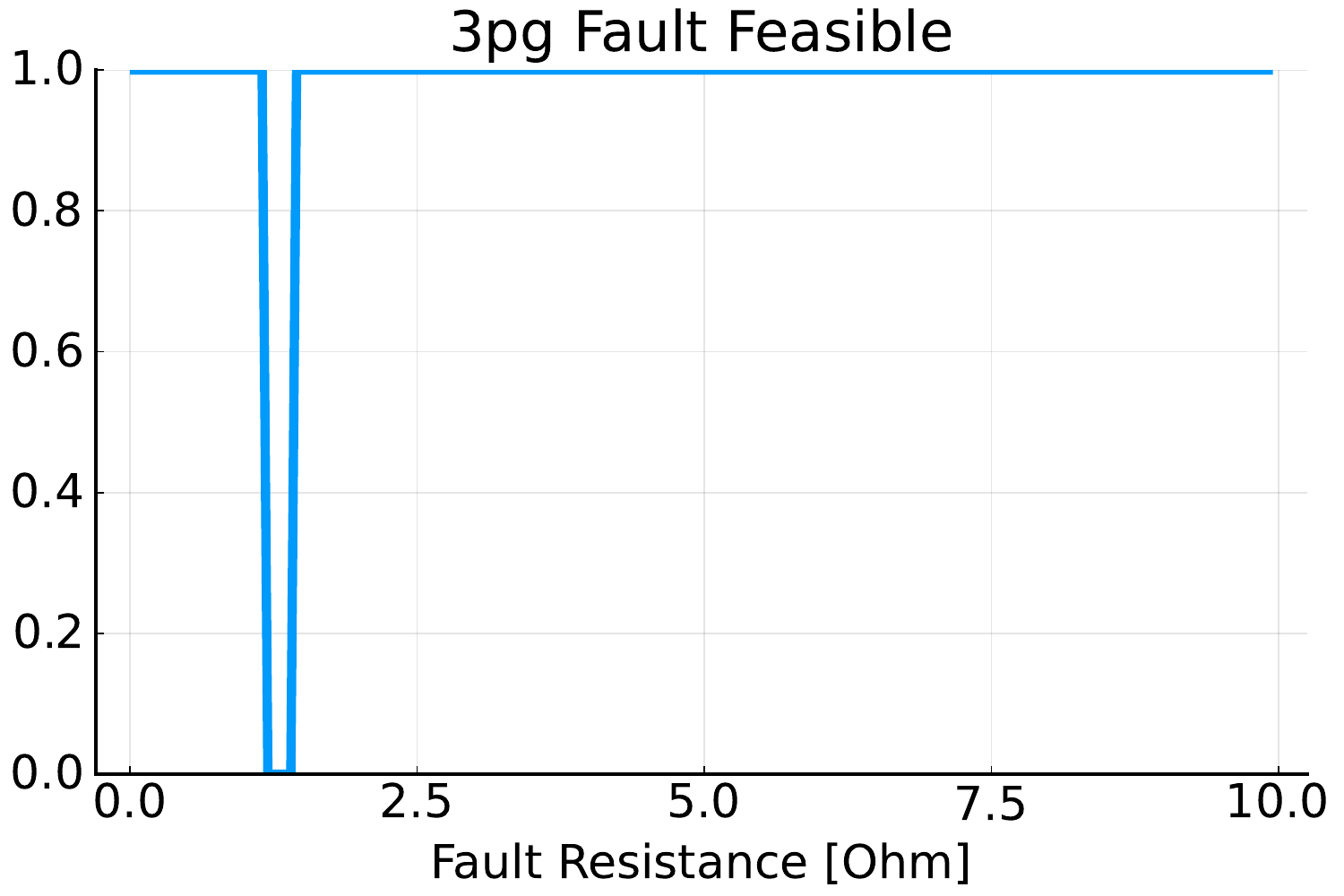}
    }
    \caption{Complex Grid-Forming Fault Currents [$A$] For 3-Phase Fault.}
    \label{fig:grid-forming-3ph_2}
\end{figure}

\begin{figure}[!htbp]
    \subfigure{
    \includegraphics[width=0.46\linewidth]{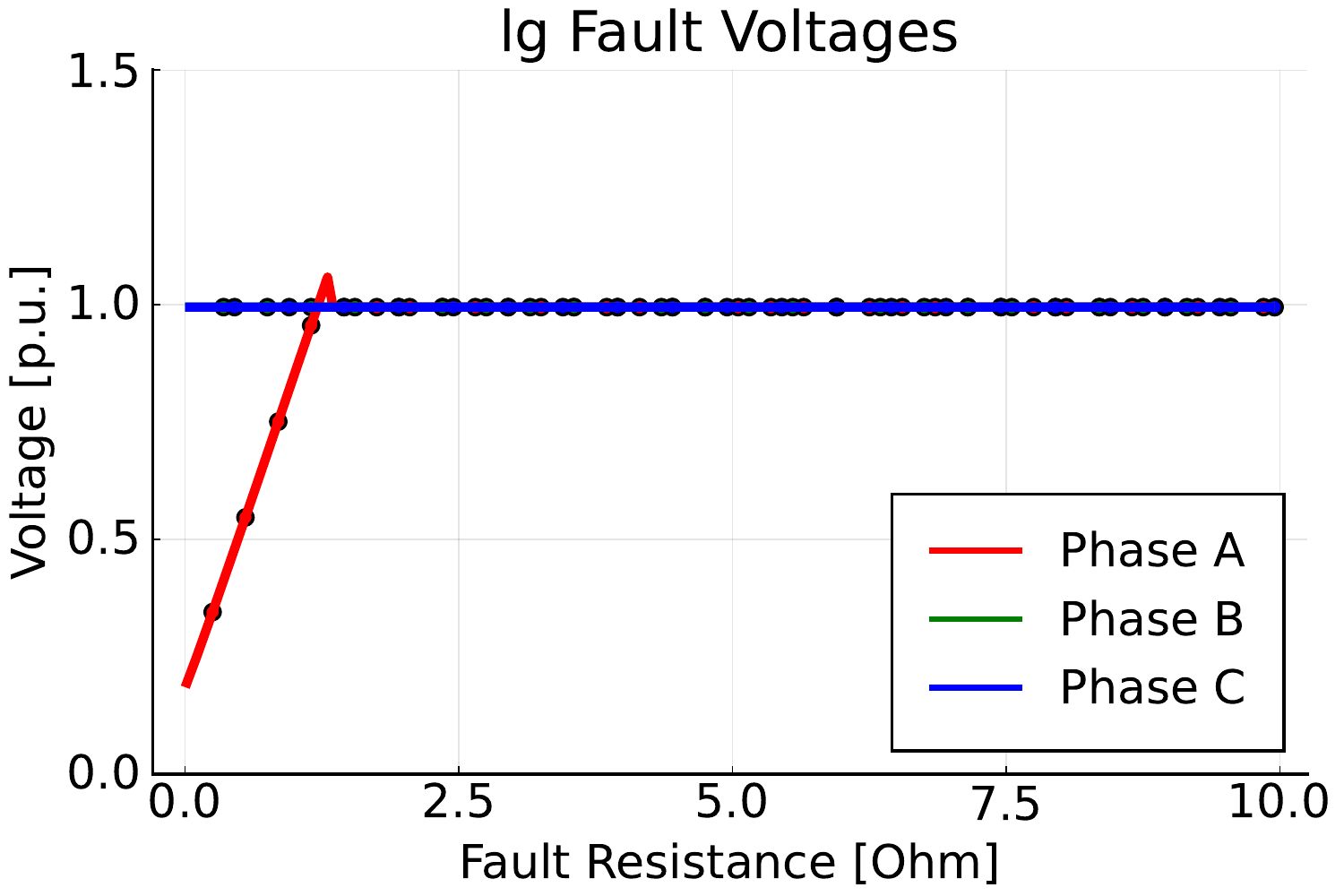}
    }
    \subfigure{
    \includegraphics[width=0.46\linewidth]{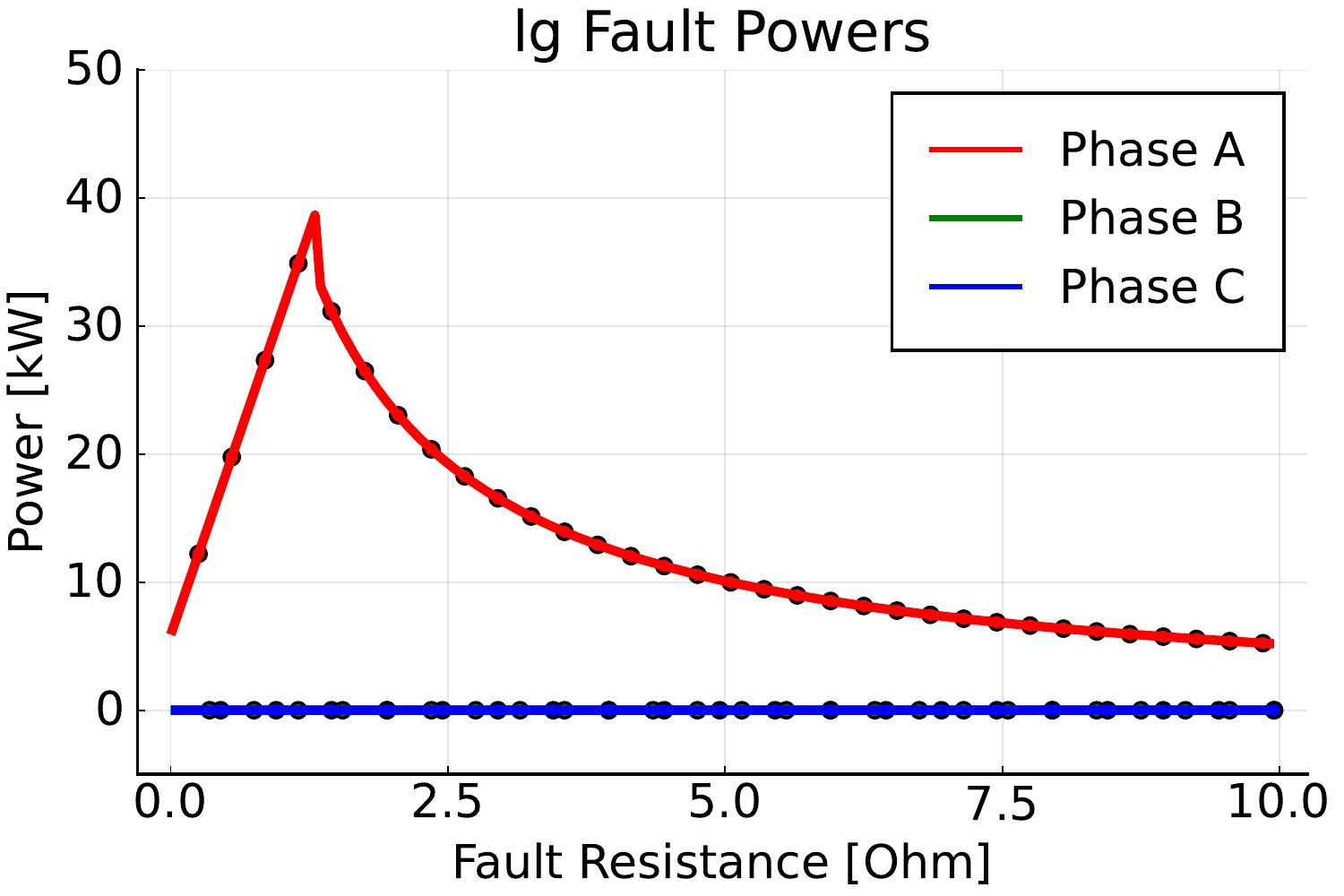}
    }
    \subfigure{
    \includegraphics[width=0.46\linewidth]{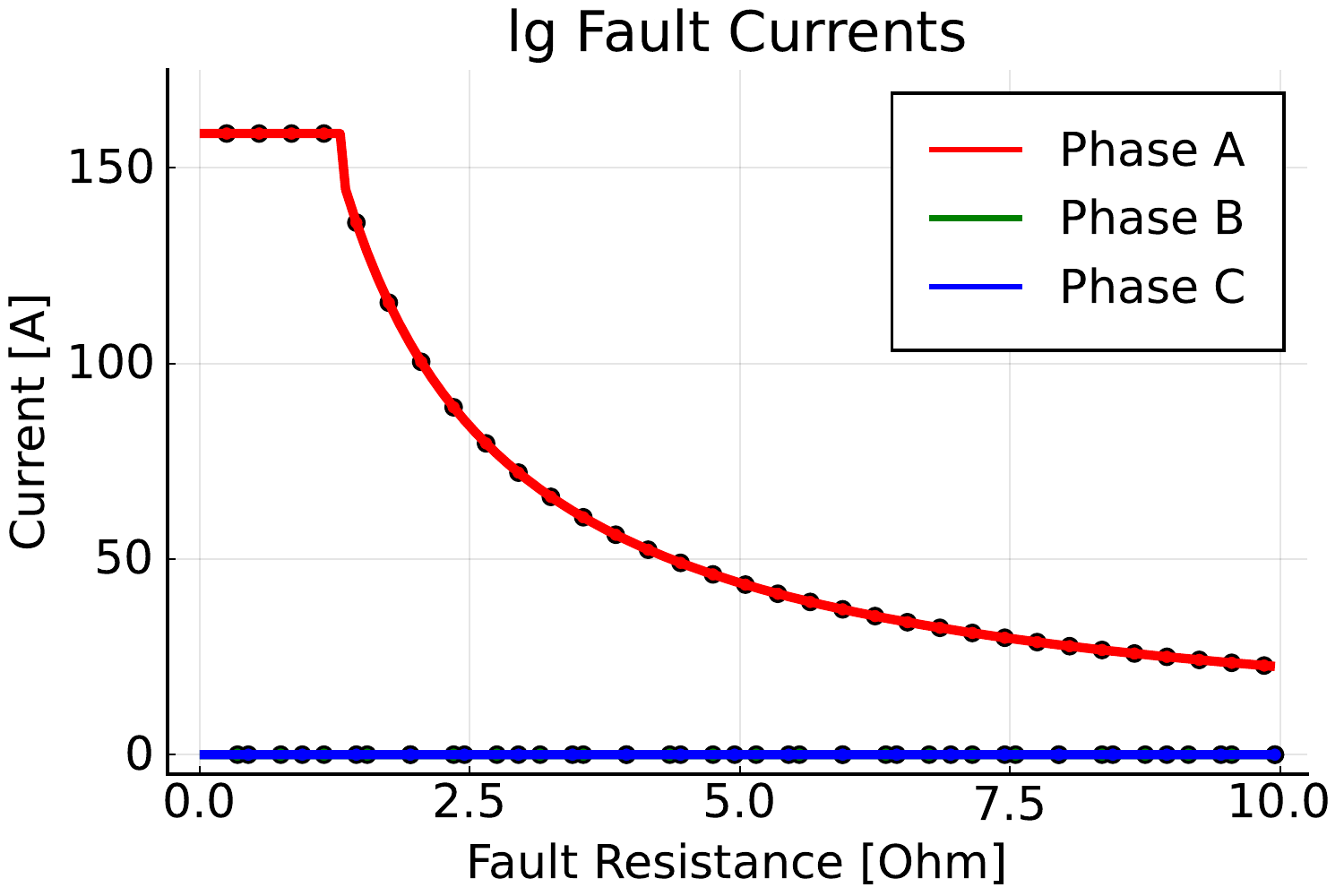}
    }
     \subfigure{
    \includegraphics[width=0.46\linewidth]{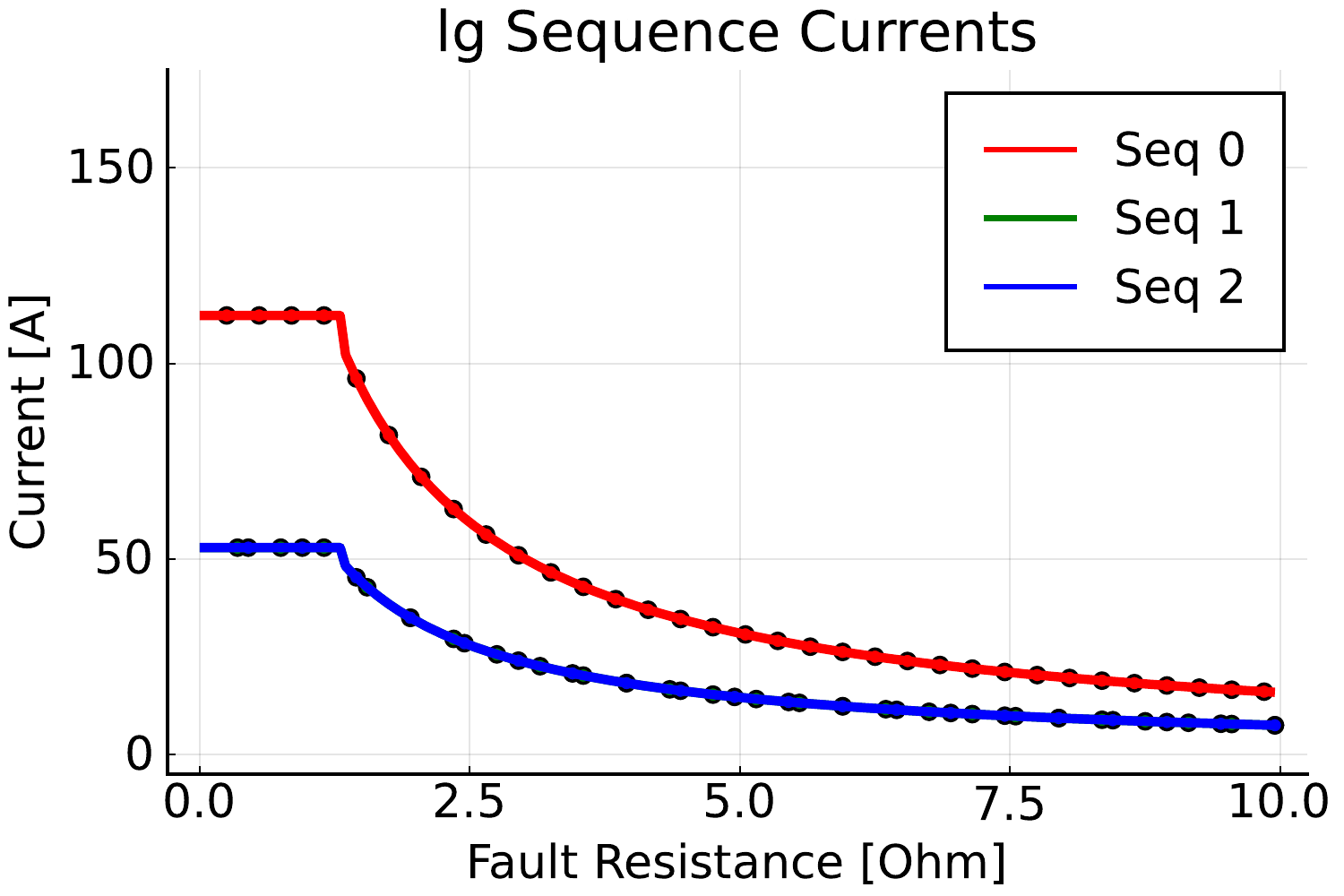}
    }
    \caption{Complex Grid-Forming Fault Currents [$A$] For Line-Ground Fault.}
    \label{fig:grid-forming-line-ground_2}
\end{figure}

The performance of the grid-forming inverter models under line-to-line, single line-to-ground, and 3-phase faults were evaluated; the results are presented in Fig.~\ref{fig:grid-forming-line-line}–\ref{fig:grid-forming-line-ground_2}.

The voltage-current relationship for the grid-forming inverters during and outside saturation match the relationship presented in \cite{8980892}. 
The experimental results presented in \cite{9254562} showed the sequence currents supplied by the grid-forming inverters for various fault types, but provided little to no information about the injected powers during faulted conditions. 
Based on the sequence current results, the developed simple and complex grid-forming inverter models appear to provide comparable sequence current injections.
However, it can be observed that unlike the developed grid-following inverter model, the grid-forming models fail to be feasible for certain fault types.

The simple grid-forming inverter model failed to provide feasible solutions for all fault impedances in the line-line fault (Fig.~\ref{fig:grid-forming-line-line}).
This likely occurred because the source was too stiff, as in \cite{9254562} it was observed that the angles of the faulted phases jumped by 40 degrees.
This model was incapable of allowing the voltage source to adjust enough to handle the fault, which can be seen in the jumps in voltage in Fig.~\ref{fig:grid-forming-line-line}.

The complex grid-forming inverter model failed to provide feasible solutions for all fault impedances in the 3-phase fault (Fig.~\ref{fig:grid-forming-3ph_2}).
From the results, the voltages and currents do not seem to exhibit any harsh jumps and do not seem to indicate an obvious issue resulting from the voltage or current constraints. It appears that the power constraints are the source of infeasibility, which can be seen by comparing Fig. \ref{fig:grid-forming-3ph} and Fig. \ref{fig:grid-forming-3ph_2}.

It must be noted that the voltages in the line-to-line fault exhibited similar jumps as was seen in the simple grid-forming inverter model, but unlike that model, the complex model was able to allow the voltage magnitudes to decrease more to allow the solution to be feasible. The jumps in the voltage magnitudes can be attributed to the voltage magnitude constraint, which is applied to both faulted phases, not establishing a preference on which phase should be lower than the other during the fault, and hence both solutions are feasible.

\section{Conclusions} \label{sec:conclusions}
\normalsize \indent

The introduced continuous variable inverter models for balanced and unbalanced distribution networks are improved versions of the preliminary models used in PMsP for pf, opf, and fault studies.
The developed grid-following inverter model presented in this work is able to provide feasible and reasonable results for a variety of common fault types.
On the other hand, the developed simple and complex grid-forming inverter models still require future work to improve behavior:~to correct the infeasible results for line-to-line and 3-phase faults observed in the simple and the complex models, respectively.
From the simulation results, the modeling of the inverter as a Thevenin equivalent source did not allow the voltage angles to adjust enough to allow for a feasible solution. Usually, Thevenin sources are capable of sourcing a large amount of current to allow for the proper voltages at the terminals. This is not possible with the grid-forming inverter models because of current saturation especially at low fault impedances.
The complex model is capable of adjusting the terminal voltage of the inverter to compensate for current saturation, but has a limit on the amount of power that can be injected into the system which caused infeasibility on the transition between saturated and non-saturated current injection in the 3-phase fault.
The power limit was not an issue with the simple model because there was no constraint on power injection with a Thevenin source. Future work will look on combining the simple and complex model to overcome the limitation of the two models.   

This work has laid down the foundation for developing relaxed models for both grid-following and grid-forming inverter models for use in optimization problems.
The use of continuous activation variables (to indicate inverter current saturation) over the use of mixed-integer formulation will allow for the inverter models to be integrated into optimization problems on larger, more complex distribution networks.



\vspace{0.25in}
\section{Acknowledgement}
\normalsize \indent

This work was performed with the support of the U.S. Department of Energy (DOE) Office of Electricity (OE) Microgrid Research and Development (MRD) Research Program under program manager Dan Ton. We gratefully acknowledge Dan's support of this work.
The research work conducted at Los Alamos National Laboratory is done under the auspices of the National Nuclear Security Administration (NNSA) of the U.S. Department of Energy (OE) under Contract No. 89233218CNA000001.
Approved for public release; distribution is unlimited -- LA-UR-22-32454.


\bibliographystyle{unsrt}
\bibliography{references}

\begin{thebibliography}{10}

\bibitem{5275777}
J.~A.~{Martinez} et~al.
\newblock {Impact of Distributed Generation on Distribution Protection and
  Power Quality}.
\newblock In {\em {Proc. of the 2009 IEEE Power \& Energy Society General
  Meeting}}, pages 1--6, Jul. 2009.

\bibitem{inverImp}
M.~{Meskin} et~al.
\newblock {Impact of Distributed Generation on the Protection Systems of
  Distribution Networks}.
\newblock {\em {IET Generation, Transmission \& Distribution}},
  14(24):5944--5960, Nov. 2020.

\bibitem{8274697}
A.~{Mishra} et~al.
\newblock {Fault Current Characterisation of Single Phase Inverter Systems}.
\newblock In {\em {Proc. of the 2017 IEEE Power \& Energy Society General
  Meeting}}, pages 1--5, 2017.

\bibitem{8669457}
D.~{Duckwitz} et~al.
\newblock {Experimental Short-Circuit Testing of Grid-Forming Inverters in
  Microgrid and Interconnected Mode}.
\newblock In {\em {Proc. of the 2018 Conference on Sustainable Energy Supply
  and Energy Storage Systems}}, pages 1--6, 2018.

\bibitem{8673877}
G.~{Kou} et~al.
\newblock {Fault Characteristics of Distributed Solar Generation}.
\newblock {\em {IEEE Transactions on Power Delivery}}, 35(2):1062--1064, Apr.
  2020.

\bibitem{9254562}
N.~S.~{Gurule} et~al.
\newblock {Experimental Evaluation of Grid-Forming Inverters Under Unbalanced
  and Fault Conditions}.
\newblock In {\em {Proc. of the 46th Annual Conference of the IEEE Industrial
  Electronics Society}}, pages 4057--4062, 2020.

\bibitem{8980892}
N.~S.~{Gurule} et~al.
\newblock {Grid-forming Inverter Experimental Testing of Fault Current
  Contributions}.
\newblock In {\em {Proc. of the 2019 IEEE 46th Photovoltaic Specialists
  Conference}}, pages 3150--3155, 2019.

\bibitem{8547488}
J.~{Hernandez-Alvidrez} et~al.
\newblock {PV-Inverter Dynamic Model Validation and Comparison Under Fault
  Scenarios Using a Power Hardware-in-the-Loop Testbed}.
\newblock In {\em {Proc. of the 2018 IEEE 7th World Conference on Photovoltaic
  Energy Conversion}}, pages 1412--1417, 2018.

\bibitem{barnes21-pmsp}
A.~K.~{Barnes} et~al.
\newblock {Optimization-Based Formulations for Short-Circuit Studies with
  Inverter-Interfaced Generation in PowerModelsProtection.jl}.
\newblock {\em MDPI Energies}, 14(8):1--27, Apr. 2021.

\bibitem{tabarez22-fccopf-micgrd}
J.~E.~{Tabarez} et~al.
\newblock {Fault Current-Constrained Optimal Power Flow on Unbalanced
  Distribution Networks}.
\newblock In {\em {Proc. of the 2022 IEEE ISGT ASIA Conference}}, pages 1--5,
  Nov. 2022.

\bibitem{PMs}
C.~{Coffrin} et~al.
\newblock {PowerModels.jl: An Open-Source Framework for Exploring Power Flow
  Formulations}.
\newblock In {\em Proc. of the 2018 Power Systems Computation Conference},
  pages 1--8, Jun. 2018.

\bibitem{PMsD}
D.~M.~{Fobes} et~al.
\newblock {PowerModelsDistribution.jl: An Open-Source Framework for Exploring
  Distribution Power Flow Formulations}.
\newblock {\em Electric Power Systems Research}, 189(106664):1--7, 2020.

\bibitem{noauthor_powermodelsprotectionjltestdatadist_nodate}
{PowerModelsProtection}.jl/test/data/dist at master ·
  lanl-ansi/{PowerModelsProtection}.jl.

\end{thebibliography}

\end{document}